\documentclass[12pt]{iopart}

\usepackage{iopams}
\usepackage{graphicx}
\usepackage{color}
\usepackage[caption=false]{subfig}
\usepackage[strings]{underscore}  
\usepackage{array}
\usepackage{textcomp}

\newcommand{\diff}{\mathrm{d}}
\newcommand{\im}{\mathrm{i}}
\newcommand{\transpose}{\top}

\begin{document}

\title{Large Deviation Analysis of Function Sensitivity in Random Deep Neural Networks}

\author{Bo Li and David Saad}
\address{Non-linearity and Complexity Research Group, Aston University, Birmingham, B4 7ET, United Kingdom}
\ead{b.li10@aston.ac.uk and d.saad@aston.ac.uk}


\begin{abstract}
Mean field theory has been successfully used to analyze deep neural networks (DNN) in the infinite size limit. Given the finite size of realistic DNN, we utilize the large deviation theory and path integral analysis to study the deviation of functions represented by DNN from their typical mean field solutions. The parameter perturbations investigated include weight sparsification (dilution) and binarization, which are commonly used in model simplification, for both ReLU and sign activation functions. We find that random networks with ReLU activation are more robust to parameter perturbations with respect to their counterparts with sign activation, which arguably is reflected in the simplicity of the functions they generate.
\end{abstract}

%
\vspace{2pc}
\noindent{\it Keywords}: large deviation theory, path integral, deep neural networks, function sensitivity
%
%
%
%

\section{Introduction}
Learning machines realized by deep neural networks (DNN) have achieved impressive success in performing various machine learning tasks, such as speech recognition, image classification and natural language processing~\cite{LeCun2015}. While DNN typically have numerous parameters and their training comes at a high computational cost, their applications have been extended also to include devices with limited memory or computational resources, such as mobile devices, thanks to compressed networks and reduced parameter precision~\cite{Cheng2018}. Most supervised learning scenarios are of DNN functions representing some input-output mapping, on the basis of input-output example patterns. DNN parameter estimation (training) aims at obtaining a network that approximates well the underlying mapping. Despite their profound engineering success, a comprehensive understanding of the intrinsic working mechanism~\cite{Zeiler2014, Yosinski2015} and the generalization ability~\cite{Chiyuan2017, Chaudhari2017, Neyshabur2017, Bartlett2017} of DNN are still lacking. The difficulty in analyzing DNN is due to the recursive nonlinear mapping between layers they implement and the coupling to data and learning dynamics. 

A recent line of research utilizes the mean field theory in statistical physics to investigate various DNN characteristics, such as expressive power~\cite{Poole2016}, Gaussian process-like behaviors of wide DNN~\cite{Duvenaud2014, Daniely2016, Lee2018}, dynamical stability in layer propagation and its impact on weight initialization~\cite{Schoenholz2017, Yang2017, Pretorius2018} and function similarity and entropy in the function space~\cite{BoLi2018}. By assuming large layer-width and random weights, such techniques harness the specific type of nonlinearity used and many degrees of freedom to provide valuable analytical insights. The Gaussian process perspectives of infinitely wide DNN also facilitates the analysis of training dynamics and generalization by employing established kernel methods~\cite{Jacot2018, Arora2019}.

To study the entropy of functions realized by DNN~\cite{BoLi2018}, we adopted similar assumptions but employed the generating functional analysis~\cite{Mozeika2009, Mozeika2010}, which is more general and can be applied to sparse and weight-correlated networks. The analysis of function error incurred by weight perturbations exhibits an exponential growth in error for DNN with sign activation functions, while networks with ReLU activation function are more robust to perturbations. We have also found that ReLU activation induces correlations among variables in random convolution networks~\cite{BoLi2018}. The robustness of random networks with ReLU activation is related to the simplicity of the functions they compute~\cite{Valle-Perez2018, Palma2019}, which may converge to a constant function in the large depth and width limit~\cite{Pretorius2018}, although, in principle, they admit high capacity with arbitrary weights. However, DNN used in practice are of finite size and finite depth, therefore it is essential to analyze the deviation of finite-size systems with respect to the typical mean field behavior, and characterize its rate of convergence with increasing size. An example of a recent study along these lines~\cite{Antognini2019} investigates the deviation in performance of finite size neural networks with a single hidden layer from the Gaussian process behavior.

In this work, we adopt the large deviation approach and the path integral formalism of~\cite{BoLi2018} to derive the deviation of function sensitivity of finite systems from their infinite system counterparts, which is applicable to a range of DNN structures. We analyze the effect of sparsifying (diluting) and binarizing DNN weights, commonly used for model simplification~\cite{LeCun1989, Hubara2016, Rastegari2016, Hou2017}. Although the dependence on data and training are not considered, the analysis of random DNN provides valuable insights and baseline comparisons. We will also investigate the sensitivity of functions to input perturbation~\cite{Poole2016, Schoenholz2017}, which is related to function complexity and generalization~\cite{Franco2006, Novak2018, Valle-Perez2018, Palma2019}. The paper is organized as follows. In Sec.~\ref{sec:model} and~\ref{sec:GF}, we introduce the random DNN model and review the basic results of generating functional analysis,respectively. In Sec.~\ref{sec:LDP_parameter} and~\ref{sec:LDP_input}, we derive the large deviation of function sensitivity to weight and input perturbations, respectively, based on the path integral formalism. Finally, in Sec.~\ref{sec:result}, we discuss the results and their implications.

\section{The model} \label{sec:model}
\begin{figure}
    \centering
    \includegraphics[scale=0.5]{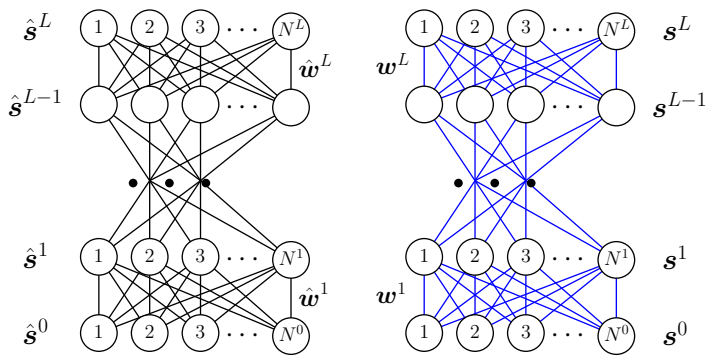}
    \caption{The reference and perturbed fully-connected DNN, parameterized by $\{\hat{\boldsymbol{w}}^{l}\}$ (black edges) and $\{\boldsymbol{w}^{l}\}$ (blue edges), respectively. Each layer $l$ has $N^{l} = \alpha^{l} N$ nodes.}
    \label{fig:two_systems}
\end{figure}

Following~\cite{BoLi2018}, we consider two coupled fully-connected DNN. One of them serves as the reference function under consideration, and the other as its perturbed counterpart, either in the weights or input variables. As shown in Fig.~\ref{fig:two_systems}, each network consists of $L+1$ layer; layer $l$ has $N^{l}$ neurons, which can be layer dependent. The reference network is parameterized by the weight variables\footnote{The usual bias variables are omitted for simplicity, but it can be easily accommodated within the current framework.} $\{ \hat{\boldsymbol{w}}^{l} \}_{l=1}^{L} $, while the perturbed network is parameterized with $\{ \boldsymbol{w}^{l} \}_{l=1}^{L} $. Similarly, variables with a circumflex are associated with the reference network. In the following, $\boldsymbol{w}^{l}$ represents the $N^{l}\times N^{l-1}$ weight matrix at layer $l$, and $\boldsymbol{w}^{l}_{i}$ represents the $N^{l-1}$ dimensional weight vector of the $i$th perceptron at layer $l$. Denoting the input dimension as $N=N^{0}$, we assume the sizes of all layers scale linearly with $N$ as $N^{l} = \alpha^{l} N$.

A deterministic feed-forward network is defined by the recursive mapping $\forall \; 1\leq l \leq L$
\begin{eqnarray}
    h^{l}_{i} = \frac{1}{\sqrt{N^{l-1}}} \sum_{j=1}^{N^{l-1}} w^{l}_{ij} s^{l-1}_{j}, \label{eq:def_hli} \\
    s^{l}_{i} = \phi^{l} \big( h^{l}_i \big), \label{eq:def_sli}
\end{eqnarray}
where $\{ w^{l}_{ij} \}$ are the weights, $h^{l}_{i}$ and $s^{l}_{i}$ are pre- and post-activation field and variable, respectively, and $\phi^{l}(\cdot)$ is the activation/transfer function at layer $l$. The scaling factor of $1/\sqrt{N^{l-1}}$ in Eq.~(\ref{eq:def_hli}) is introduced for normalization. We primarily focus on networks with either sign $\left[\phi_{s}(x) = \mathrm{sgn}(x)\right]$ or ReLU $\left[\phi_{r}(x) = \max(x, 0)\right]$ activation functions in the hidden layers, and consider binary input and output variables $s^{0}_{i}, s^{L}_{i} \in \{ 1, -1 \}$ by applying the sign activation function at the output layer $s^{L}_{i} = \mathrm{sgn}(h^{L}_{i})$ for a fair comparison across architectures. The resulting feed-forward DNN implements a Boolean mapping $f:\{1,-1\}^{N^{0}} \to \{1,-1\}^{N^{L}}$, where each output node $s^{L}_{i}\big( \boldsymbol{s}^{0} \big)$ computes a Boolean function. In the following, we call the two architectures sign-DNN and relu-DNN respectively, keeping in mind that sign activation function is always applied in the output layer.

To facilitate a path integral calculation, we consider stochastic dynamics between successive layers. For the layer with sign activation function, the activation $s^{l}_{i}$ is disturbed by thermal noise according to the following probability
\begin{equation}
    P\big(s^{l}_{i} | h^{l}_{i}(\boldsymbol{w}^{l}, \boldsymbol{s}^{l-1}) \big) = \frac{ \exp \big( \beta s^{l}_{i} h^{l}_{i}(\boldsymbol{w}^{l}, \boldsymbol{s}^{l-1}) \big) }{ 2 \cosh \big( \beta h^{l}_{i}(\boldsymbol{w}^{l}, \boldsymbol{s}^{l-1}) \big) },
\end{equation}
while for relu activation function, $s^{l}_{i}$ is disturbed by additive Gaussian noise
\begin{equation}
    P\big(s^{l}_{i} | h^{l}_{i}(\boldsymbol{w}^{l}, \boldsymbol{s}^{l-1}) \big) = \sqrt{\frac{\beta}{2\pi}} \exp \bigg\{ -\frac{\beta}{2} \bigg[ s^{l}_{i} - \phi\big( h^{l}_{i}(\boldsymbol{w}^{l}, \boldsymbol{s}^{l-1}) \big) \bigg]^2 \bigg\}.
\end{equation}
In the limit $\beta \to \infty$, we recover the deterministic model. The evolution of the two systems follows the joint distribution
\begin{equation}
    P(\{ \hat{s}^{l}_{i}, s^{l}_{i} \}) = P(\hat{\boldsymbol{s}}^{0}, \boldsymbol{s}^{0}) \prod_{l=1}^{L} \prod_{i=1}^{N^{l}} P\big(\hat{s}^{l}_{i} | \hat{h}^{l}_{i}(\hat{\boldsymbol{w}}^{l}, \hat{\boldsymbol{s}}^{l-1}) \big) P\big(s^{l}_{i} | h^{l}_{i}(\boldsymbol{w}^{l}, \boldsymbol{s}^{l-1}) \big). \label{eq:joint_prob_shat_s}
\end{equation}

\begin{figure}
	\centering
	\includegraphics[scale=0.5]{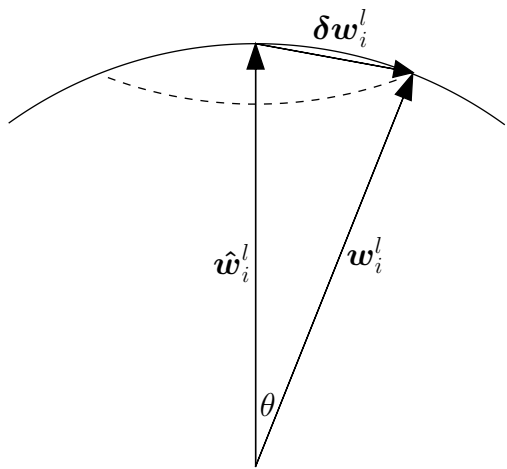}
	\caption{A geometric representation of perturbations on the parameter vector $\hat{\boldsymbol{w}}^{l}_{i}$ defined in Eq.~\ref{eq:def_eta_perturbation}, resulting in a rotated vector $\boldsymbol{w}^{l}_{i}$ at an angle $\theta^{l} = \sin^{-1}\eta^{l}$.}
	\label{fig:NoiseOnSphere}
\end{figure}

To probe the difference between the functions implemented by the two networks, we feed in the same \textit{single} input $\boldsymbol{s}^{0} = \hat{\boldsymbol{s}}^{0}$ to the two systems such that $P(\hat{\boldsymbol{s}}^{0}, \boldsymbol{s}^{0}) =  P(\hat{\boldsymbol{s}}^{0}) \prod_{i=1}^{N^{0}} \delta_{\hat{s}^{0}_{i}, s^{0}_{i}}$, and study the resulting output difference due to parameter perturbation. For continuous weight variables, one useful choice for the weight perturbation is
\begin{equation}
    w^{l}_{ij} = \sqrt{1 - (\eta^{l})^2} \hat{w}^{l}_{ij} + \eta^{l} \delta w^{l}_{ij}, \label{eq:def_eta_perturbation} 
\end{equation}
which ensures that $w^{l}_{ij}$ has the same variance of $\hat{w}^{l}_{ij}$ as long as $\delta w^{l}_{ij}$ follows the same distribution of $\hat{w}^{l}_{ij}$, and effectively rotates the high dimensional vector $\hat{\boldsymbol{w}}^{l}_{i}$ by an angle $\theta^{l} = \sin^{-1}\eta^{l}$ as demonstrated schematically in Fig.~\ref{fig:NoiseOnSphere}. 

In probing the sensitivity of a function due to input perturbations, the weights of two networks are kept the same $\boldsymbol{w} = \hat{\boldsymbol{w}}$ and a fixed fraction of input variables are flipped randomly. The resulting output difference of the two systems reflects the sensitivity and complexity of the underlying DNN.

\section{Generating functional analysis for typical behavior} \label{sec:GF}
Viewing the weights $\{ \hat{w}^{l}_{ij}, w^{l}_{ij} \}$ as quenched random variables, a generating functional analysis has been proposed~\cite{BoLi2018} to derive the typical behavior of DNN. It starts with computing the disorder-averaged generating functional 
\begin{equation}
    \overline{\Gamma}(\hat{\boldsymbol{\psi}}, \boldsymbol{\psi}) = \mathbb{E}_{\hat{\boldsymbol{w}}, \boldsymbol{w}} \mathbb{E}_{\hat{\boldsymbol{s}}, \boldsymbol{s} } \exp \bigg( - \im \sum_{l,i} ( \hat{\psi}^{l}_{i} \hat{s}^{l}_{i} + \psi^{l}_{i} s^{l}_{i} )  \bigg),
\end{equation}
where the average $\mathbb{E}_{\hat{\boldsymbol{s}}, \boldsymbol{s} }$ is taken with respect to the joint probability Eq.~(\ref{eq:joint_prob_shat_s}). Assume the layer widths are the same $N^{l}=N$ for all $l$. Upon averaging over the disorder $\hat{\boldsymbol{w}}, \boldsymbol{w}$, the generating functional can be expressed through a set of macroscopic order parameters such as the overlaps $q^{l} = 1/N^{l} \sum_i \langle \hat{s}^{l}_{i} s^{l}_{i} \rangle$ and magnetizations $\hat{m}^{l} = 1/N^{l} \sum_i \langle \hat{s}^{l}_{i} \rangle, m^{l} = 1/N^{l} \sum_i \langle s^{l}_{i} \rangle$ as
\begin{equation}
    \overline{\Gamma} = \int \{ \diff\boldsymbol{q} \diff\boldsymbol{Q} ... \} \exp\big[ N\Psi(\boldsymbol{q}, \boldsymbol{Q}, ...) \big].
\end{equation}
where $\boldsymbol{Q}$ is the conjugate variable of the order parameter $\boldsymbol{q}$. In the large system size limit $N \to \infty$, the generating functional  $\overline{\Gamma}$ is dominated by the saddle point of the potential function $\Psi(\boldsymbol{q}, \boldsymbol{Q}, ...)$. It gives rise to typical overlaps that dominate in probability, which facilitates analytical studies of random DNN.

Assume the weight perturbation follows the form of Eq.~(\ref{eq:def_eta_perturbation}), and both weight and perturbation are independent of each other and follow a Gaussian distribution $\hat{w}^{l}_{ij}, \delta w^{l}_{ij} \sim \mathcal{N}(0,\sigma_{w}^{2})$. It is found that for the layer with sign activation function in the limit $\beta \to \infty$, the overlap evolves as~\cite{BoLi2018}
\begin{equation}
    q^{l}=\frac{2}{\pi}\sin^{-1}\left(\sqrt{1-(\eta^{l})^2}q^{l-1}\right), \quad 1 \leq l \leq L. \label{eq:mf_propagation_sign}
\end{equation}
Similarly, for ReLU activation function in the deterministic limit, if the weight standard deviation is chosen as $\sigma_{w}=\sqrt{2}$, the magnitude of the activations remains stable and the overlap evolves as
\begin{eqnarray}
    q^{l} = & \frac{1}{\pi} \bigg\{ \sqrt{1- \big[ 1-(\eta^{l})^2 \big] (q^{l-1})^2 } \nonumber \\
    \quad & + \sqrt{1-(\eta^{l})^2}q^{l-1} \bigg[ \frac{\pi}{2} + \sin^{-1}\left(\sqrt{1-(\eta^{l})^2}q^{l-1}\right) \bigg] \bigg\}, \label{eq:mf_propagation_relu}
\end{eqnarray}
while the output layer $L$ follows Eq.~(\ref{eq:mf_propagation_sign}) due to the use of the sign activation function. The restriction $\boldsymbol{s}^{0} = \hat{\boldsymbol{s}}^{0}$ leads to $q^{0} = 1$ in both cases.

\section{Large deviations in parameter sensitivity of functions} \label{sec:LDP_parameter}
The generating functional analysis above gives typical behaviors of random DNN in the limit $N \to \infty$. However, practical DNN always have finite sizes. Therefore, it is worthwhile to understand the deviation to the most probable behaviors under finite $N$. In the following, we adopt the large deviation analysis to tackle this problem. An introduction of large deviation theory and its application to statistical mechanics can be found in~\cite{Touchette2009}. In essence, a continuous observable $\mathcal{O}$ in a system of size $N$ (assumed to be large) is said to satisfy the large deviation principle if the probability of finding  $\mathcal{O}$ follows
\begin{equation}
    \mathrm{Prob}_{N}(\mathcal{O} \in [x, x+\diff x]) \simeq e^{-N I(x)} \diff x,
\end{equation}
where $I(x)$ is the rate function of the observable. It implies that the probability density of $\mathcal{O}$ scales as $P_{N}(\mathcal{O}=x) \simeq e^{-N I(x)}$, which is concentrated at the minimum of the rate function $x^* = \mathrm{argmin}_{x} I(x)$ in large systems and the profile of $I(x)$ quantifies the fluctuation of the observable.

In this work the overlap of the output layer $q^{L}:= 1/N^{L} \sum_{i} \hat{s}^{L}_{i} s^{L}_{i}$ is at the focus of our study. The path integral techniques adopted in the generating functional  framework~\cite{BoLi2018} can be adapted to tackle the large deviation analysis. We start with computing the probability density \footnote{Here we assume $q^{L}=1/N^{L}\sum_{i=1}^{N^{L}} \hat{s}^{L}_{i} s^{L}_{i}$ to be a continuous variable by considering large $N^{L}$. Instead, one can view $q^{L}$ as a discrete variable by definition (since the inputs are binary variables), where $\delta(\cdot)$ should be understood as the Kronecker delta function.}
\begin{eqnarray}
    \fl P({q^L}) = \left\langle \delta \bigg( \frac{1}{N^{L}} \sum_{i} \hat{s}^{L}_{i} s^{L}_{i} - q^L \bigg) \right\rangle \nonumber \\
    \fl = \mathbb{E}_{\hat{\boldsymbol{w}}, \boldsymbol{w}} \mathrm{Tr}_{\boldsymbol{\hat{s}}, \boldsymbol{s}} P(\hat{\boldsymbol{s}}^{0}) \prod_{i=1}^{N^{0}}\delta_{s_{i}^{0},\hat{s}_{i}^{0}}\prod_{l=1}^{L} P(\hat{\boldsymbol{s}}^{l}|\hat{\boldsymbol{w}}^{l},\hat{\boldsymbol{s}}^{l-1}) P(\boldsymbol{s}^{l}|\boldsymbol{w}^{l},\boldsymbol{s}^{l-1})  \delta \bigg( \frac{1}{N^{L}} \sum_{i} \hat{s}^{L}_{i} s^{L}_{i} - q^L \bigg ),
\end{eqnarray}
where the operation $\mathrm{Tr}_{\hat{\boldsymbol{s}}, \boldsymbol{s}}$ is understood as an integration or summation depending on the nature of variables.
The input distribution follows $P(\hat{\boldsymbol{s}}^{0})=\prod_{i}P(\hat{s}_{i}^{0})=\prod_{i}(\frac{1}{2}\delta_{\hat{s}_{i}^{0},1}+\frac{1}{2}\delta_{\hat{s}_{i}^{0},-1})$. To deal with the non-linearity of the pre-activation fields in the conditional probability, we introduce auxiliary fields $\{ \hat{x}^{l}_{i}, x^{l}_{i} \}$ through
the integral representation of delta-function
\begin{equation}
    \fl 1 = \int_{-\infty}^{\infty}\frac{\diff\hat{h}_{i}^{l} \diff\hat{x}_{i}^{l}}{2\pi} e^{\im\hat{x}_{i}^{l}\left(\hat{h}_{i}^{l}-\frac{1}{\sqrt{N^{l-1}}}\sum_{j}\hat{w}_{ij}^{l}\hat{s}_{j}^{l-1}\right)}, \quad 1 = \int_{-\infty}^{\infty}\frac{\diff h_{i}^{l} \diff x_{i}^{l}}{2\pi}e^{\im x_{i}^{l}\left(h_{i}^{l}-\frac{1}{\sqrt{N^{l-1}}}\sum_{j}w_{ij}^{l}s_{j}^{l-1}\right)},
\end{equation}
which allows us to express the quenched random variables $\hat{w}_{ij}^{l}$
and $w_{ij}^{l}$ linearly in the exponents, leading to
\begin{eqnarray}
    \fl P(q^{L}) & = \mathbb{E}_{\hat{\boldsymbol{w}}, \boldsymbol{w}} \mathrm{Tr}_{\hat{\boldsymbol{s}}, \boldsymbol{s}} \delta\left(\frac{1}{N^{L}}\hat{s}^{L}_{i} s^{L}_{i} - q^{L}\right) \prod_{i=1}^{N^{0}} P(\hat{s}^{0}_{i}) \delta_{s_{i}^{0},\hat{s}_{i}^{0}} \int \prod_{l=1}^{L} \prod_{i=1}^{N^{l}} \frac{\diff \hat{h}^{l}_{i} \diff \hat{x}^{l}_{i} }{2\pi} \frac{\diff h^{l}_{i} \diff x^{l}_{i} }{2\pi} \nonumber \\
    \fl & \quad \times \exp \left[ \sum_{l=1}^{L} \sum_{i=1}^{N^{l}} \bigg( \log P(\hat{s}^{l}_{i}|\hat{h}^{l}_{i}) + \log P(s^{l}_{i}|h^{l}_{i}) + \im \hat{x}^{l}_{i} \hat{h}^{l}_{i} + \im x^{l}_{i} h^{l}_{i} \bigg) \right]  \nonumber \\
    \fl & \quad \times \exp \left[ - \sum_{l=1}^{L} \frac{\im}{\sqrt{N^{l-1}}} \sum_{i=1}^{N^{l}} \sum_{j=1}^{N^{l-1}} \bigg( \hat{w}^{l}_{ij} \hat{x}^{l}_{i} \hat{s}^{l-1}_{j} + w^{l}_{ij} x^{l}_{i} s^{l-1}_{j} \bigg) \right]. \label{eq:PqL_before_average}
\end{eqnarray}
Assuming self-averaging~\cite{Dominicis1978} we exchange the order of summation and integration, and first carry out the average over the disorder variables. Specifically, we consider the weights of the reference network to be independent and follow a Gaussian distribution $\hat{w}^{l}_{ij} \sim \mathcal{N}(0, \sigma_{w}^2)$ as before, and three types of perturbations
\begin{enumerate}
    \item rotation of the weight vector $\hat{\boldsymbol{w}}^{l}_{i}$ following Eq.~(\ref{eq:def_eta_perturbation});
    \item sparsification of the weight matrix $\hat{\boldsymbol{w}}^{l}$ by randomly dropping connections with probability $p^{l}$ and rescaling the remaining weights by $1/\sqrt{1-p^{l}}$ to ensure the same weight strength
    \begin{equation}
        w^{l}_{ij} = \left\{ \begin{array}{ccc}
        0, & \mbox{with probability } p^{l}, \\
        \frac{1}{\sqrt{1-p^{l}}}\hat{w}^{l}_{ij}, & \mbox{ with probability } 1-p^{l},
        \end{array}
        \right.
        \label{eq:def_sparse_perturbation}
    \end{equation}
    \item binarization of weight element $\hat{w}^{l}_{ij}$
    \begin{equation}
        w^{l}_{ij} = \mathrm{sgn}(\hat{w}^{l}_{ij}) \sigma_{w}, \label{eq:def_binary_perturbation}
    \end{equation}
    where $\sigma_{w}$ is introduced for keeping the variance of $w^{l}_{ij}$ the same as $\hat{w}^{l}_{ij}$.
\end{enumerate}

\subsection{Macroscopic order parameters}\label{sec:macro_order_parameters}
For perturbation of type (i), the disorder average of the third line of Eq.~(\ref{eq:PqL_before_average}) yields
\begin{equation}
    \fl \prod_{l,i} \exp \left\{ \!-\!\sigma_{w}^{2} \Bigg[\frac{1}{2}(\hat{x}_{i}^{l})^{2} \frac{\sum_{j} (\hat{s}^{l-1}_{j})^2}{N^{l-1}} \!+\!\frac{1}{2}(x_{i}^{l})^{2} \frac{\sum_{j} (s^{l-1}_{j})^2}{N^{l-1}} \!+\!\sqrt{1-(\eta^{l})^2}\hat{x}_{i}^{l}x_{i}^{l} \frac{\sum_{j} \hat{s}^{l-1}_{j} s^{l-1}_{j}}{N^{l-1}} \Bigg] \right\},
     \label{eq:disorder_average_x_eta}
\end{equation}
To decouple Eqs.~(\ref{eq:PqL_before_average}) and (\ref{eq:disorder_average_x_eta}) over sites we introduce three sets of order parameters by inserting the identity
\begin{eqnarray}
    1 = \int\frac{\diff \hat{V}^{l} \diff \hat{v}^{l}}{2\pi/N^{l}}e^{\im N^{l} \hat{V}^{l}\left[\hat{v}^{l}-\frac{1}{N^{l}}\sum_{j} (\hat{s}_{j}^{l})^2 \right]}, \quad 1 = \int\frac{\diff V^{l} \diff v^{l}}{2\pi/N^{l}}e^{\im N^{l}V^{l}\left[v^{l}-\frac{1}{N^{l}}\sum_{j}(s_{j}^{l})^2\right]},\nonumber \\
    1 = \int\frac{\diff Q^{l} \diff q^{l}}{2\pi/N^{l}}e^{\im N^{l}Q^{l}\left[q^{l}-\frac{1}{N^{l}}\sum_{j} \hat{s}_{j}^{l} s_{j}^{l}\right]}, \quad \forall \; l \neq L,
\end{eqnarray}
and by expressing the output constraint as
\begin{equation}
    \delta\bigg( \frac{1}{N^{L}} \sum_{i=1}^{N^{L}} \hat{s}^{L}_{i} s^{L}_{i} - q^{L} \bigg) = \int\frac{\diff Q^{L}}{2\pi/N^{L}}e^{\im N^{L}Q^{L}\left[q^{L}-\frac{1}{N^{L}}\sum_{j} \hat{s}_{j}^{L} s_{j}^{L}\right]}.
\end{equation}

Upon introducing these macroscopic order parameters, Eq.~(\ref{eq:disorder_average_x_eta}) becomes $\prod_{l,i}\exp\{ - 1/2 [\hat{x}^{l}_{i}, x^{l}_{i}] \cdot \Sigma_{l} \cdot [\hat{x}^{l}_{i}, x^{l}_{i}]^{\transpose} \}$ with the covariance matrix $\Sigma_{l}$ 
\begin{equation}
    \Sigma_{l} := \sigma_{w}^{2} \left[ \matrix{ \hat{v}^{l-1} & \sqrt{1-(\eta^{l})^{2}}q^{l-1} \cr \sqrt{1-(\eta^{l})^{2}}q^{l-1} & v^{l-1} } \right].
\end{equation}
The probability density in Eq.~(\ref{eq:PqL_before_average}) involves $N^{l}$ identical integration and summation at each layer $l$, which can be performed individually~\cite{BoLi2018}, yielding
\begin{eqnarray}
    \fl P(q^{L}) = & \int \frac{\diff Q^{L}}{2\pi/N^{L}} \prod_{l=0}^{L-1} \frac{\diff \hat{V}^{l} \diff \hat{v}^{l}}{2\pi/N^{l}} \frac{\diff V^{l} \diff v^{l}}{2\pi/N^{l}} \frac{\diff Q^{l} \diff q^{l}}{2\pi/N^{l}} \nonumber \\
    \fl & \times e^{ \sum_{l=0}^{L-1} N^{l} \big( \im \hat{V}^{l} \hat{v}^{l} + \im V^{l} v^{l} + \im Q^{l} q^{l} \big) + N^{L} \im Q^{L} q^{L} } e^{- N^{0} \big( \im \hat{V}^{0} + \im V^{0} + \im Q^{0}  \big)} \nonumber \\
    \fl & \times \prod_{l=1}^{L-1} \left[ \int \diff H^{l} \frac{ e^{ -\frac{1}{2} (H^{l})^{\transpose} \Sigma_{l}^{-1} H^{l} } }{\sqrt{(2\pi)^2 |\Sigma_{l}}|} \mathrm{Tr}_{\hat{s}^{l}, s^{l}} P(\hat{s}^{l}|\hat{h}^{l}) P(s^{l}|h^{l}) e^{-\im \hat{V}^{l} (\hat{s}^{l})^2 - \im V^{l} (v^{l})^2 - \im Q^{l} \hat{s}^{l} s^{l} } \right]^{N^{l}} \nonumber \\
    \fl & \times \left[ \int \diff H^{L} \frac{ e^{ -\frac{1}{2} (H^{L})^{\transpose} \Sigma_{L}^{-1} H^{L} } }{\sqrt{(2\pi)^2 |\Sigma_{L}}|} \mathrm{Tr}_{\hat{s}^{L}, s^{L}} P(\hat{s}^{L}|\hat{h}^{L}) P(s^{L}|h^{L}) e^{-\im Q^{L} \hat{s}^{L} s^{L} } \right]^{N^{L}}, \label{eq:PqL_disorder_average}
\end{eqnarray}
where we have integrated out the auxiliary fields $\{ \hat{x}^{l}, x^{l} \}$ and introduced the field doublet $H^{l}:=[\hat{h}^{l},h^{l}]^{\transpose}$. We further write $P(q^{L})$ as
\begin{equation}
    \fl P(q^{L}) = \int \frac{\diff Q^{L}}{2\pi/N^{L}} \prod_{l=0}^{L-1} \frac{\diff \hat{V}^{l} \diff \hat{v}^{l}}{2\pi/N^{l}} \frac{\diff V^{l} \diff v^{l}}{2\pi/N^{l}} \frac{\diff Q^{l} \diff q^{l}}{2\pi/N^{l}} \exp[-N\Phi(\boldsymbol{Q},\boldsymbol{q},\hat{\boldsymbol{V}},\hat{\boldsymbol{v}},\boldsymbol{V},\boldsymbol{v}|q^{L})], \label{eq:PqL_eq_intexpNPhi}
\end{equation}
where $-N\Phi(\boldsymbol{Q},\boldsymbol{q},\hat{\boldsymbol{V}},\hat{\boldsymbol{v}},\boldsymbol{V},\boldsymbol{v}|q^{L})$ is equal to the logarithm of the integrand in Eq.~(\ref{eq:PqL_disorder_average}). Similar to the analysis in~\cite{BoLi2018}, the probability density $P(q^{L})$ is dominated by the saddle point $(\boldsymbol{Q}^*,\boldsymbol{q}^*,...)$ of the potential function $\Phi(...)$ in the large $N$ limit ($N^{l} = \alpha^{l} N$ with $\alpha^{l}$ as a constant) 
\begin{equation}
    P(q^{L}) \approx \exp[-N\Phi(\boldsymbol{Q}^*,\boldsymbol{q}^*,...|q^{L})],
\end{equation}
where $I(q^{L}) = \Phi(\boldsymbol{Q}^*,\boldsymbol{q}^*,...|q^{L})$ is the desired rate function.

While this set-up is based on computing the deviation in function similarity with a single input $q^{L} = 1/N^{L} \sum_{i} \hat{s}^{L}_{i} s^{L}_{i} $, one may argue that it requires testing on more than one input for obtaining a robust estimation, e.g.,
\begin{equation}
    \tilde{q}^{L} := \frac{1}{N^{L} M} \sum_{\mu=1}^{M} \sum_{i=1}^{N^{L}} \hat{s}^{L,\mu}_{i} s^{L,\mu}_{i},
\end{equation}
where $M$ is the number of independent patterns used. Assuming that representation of different patterns are uncorrelated, we show in~\ref{sec:appendix_multiple_input} that for small $M$, the rate function $I(\tilde{q}^{L})$ is approximately related to the single input case through a simple scaling
\begin{equation}
    I(\tilde{q}^{L}) \approx M\Phi(\boldsymbol{Q}^*,\boldsymbol{q}^*,...|\tilde{q}^{L}). \label{eq:rate_func_tilde_qL}
\end{equation}
This assumption is valid for sign-DNN but not for relu-DNN. We also confirm this scaling relation by numerical experiments (see below and in~\ref{sec:appendix_multiple_input}).

\subsection{Unifying three types of weight perturbations} \label{sec:unify_perturbation}
The other two types of perturbations can be treated similarly. For network sparsification~(\ref{eq:def_sparse_perturbation}), the disorder average of Eq.~(\ref{eq:PqL_before_average}) has the following form in the large $N^{l}$ limit (see~\ref{sec:appendix_average_x_sparse} for details)
\begin{equation}
    \fl \prod_{l,i} \exp \left\{ -\sigma_{w}^{2} \Bigg[\frac{1}{2}(\hat{x}_{i}^{l})^{2} \frac{\sum_{j} (\hat{s}^{l-1}_{j})^2}{N^{l-1}} +\frac{1}{2}(x_{i}^{l})^{2} \frac{\sum_{j} (s^{l-1}_{j})^2}{N^{l-1}} +\sqrt{1-p^{l}}\hat{x}_{i}^{l}x_{i}^{l} \frac{\sum_{j} \hat{s}^{l-1}_{j} s^{l-1}_{j}}{N^{l-1}} \Bigg] \right\},
     \label{eq:disorder_average_x_sparse}
\end{equation}
which has the same form of Eq.~(\ref{eq:disorder_average_x_eta}) when $p^{l}$ is replaced by $(\eta^{l})^2$. Introducing the same order parameters, we obtain the covariance of the fields $\hat{h}^{l}$ and $h^{l}$ in the form of
\begin{equation}
    \Sigma^{\mathrm{s}}_{l} := \sigma_{w}^{2} \left[ \matrix{ \hat{v}^{l-1} & \sqrt{1-p^{l}}q^{l-1} \cr \sqrt{1-p^{l}}q^{l-1} & v^{l-1} } \right].
\end{equation}
Hence, diluting connections with probability $p^{l}$ at layer $l$ in a random DNN corresponds to rotating each of the weight vector $\hat{\boldsymbol{w}}^{l}_{i}$ by an angle $\theta^{l} = \sin^{-1} \sqrt{p^{l}}$.

Similarly, for network binarization in Eq.~(\ref{eq:def_binary_perturbation}), the disorder average of Eq.~(\ref{eq:PqL_before_average}) yields (see~\ref{sec:appendix_average_x_binary} for details)
\begin{eqnarray}
    \fl \prod_{l,i} \exp \left\{ -\sigma_{w}^{2} \Bigg[\frac{1}{2}(\hat{x}_{i}^{l})^{2} \frac{\sum_{j} (\hat{s}^{l-1}_{j})^2}{N^{l-1}} +\frac{1}{2}(x_{i}^{l})^{2} \frac{\sum_{j} (s^{l-1}_{j})^2}{N^{l-1}} + \sqrt{\frac{2}{\pi}} \hat{x}_{i}^{l}x_{i}^{l} \frac{\sum_{j} \hat{s}^{l-1}_{j} s^{l-1}_{j}}{N^{l-1}} \Bigg] \right\},
     \label{eq:disorder_average_x_binary}
\end{eqnarray}
which corresponds to the covariance matrix of the fields $\hat{h}^{l}$ and $h^{l}$ to be in the form
\begin{equation}
    \Sigma^{\mathrm{b}}_{l} := \sigma_{w}^{2} \left[ \matrix{ \hat{v}^{l-1} & \sqrt{\frac{2}{\pi}}q^{l-1} \cr \sqrt{\frac{2}{\pi}}q^{l-1} & v^{l-1} } \right].
\end{equation}
Comparing to type (i) perturbation, one finds that binarizing weight elements in a random DNN corresponds to rotating each of the weight vectors $\hat{\boldsymbol{w}}^{l}_{i}$ by a fixed angle $\theta^{l} = \cos^{-1} \sqrt{\frac{2}{\pi}} \approx 37^{\circ}$. This phenomenon has been observed in~\cite{Anderson2018} and is linked to the practical success of binary DNN. It is argued~\cite{Anderson2018} that $37^{\circ}$ is a very small angle in high dimensional spaces where two randomly sampled vectors are typically orthogonal to each other; therefore weight binarization approximately preserves the directions of the high dimensional weight vectors, which contributes to the success of binary DNN. 

Therefore, we establish that the three types of perturbations on random DNN can be unified in the same framework developed in Sec.~\ref{sec:macro_order_parameters}.

\subsection{Saddle point equations}\label{sec:saddle_point_eq}
For networks with a generic activation function, the large deviation potential function $\Phi(...)$ can be express as
\begin{eqnarray}
    \fl \Phi = - \alpha^{0} \big[ \im \hat{V}^{0} (\hat{v}^{0}-1) + \im V^{0} (v^{0}-1) +  \im Q^{0} (q^{0}-1) \big] -\sum_{l=1}^{L-1} \alpha^{l} (\im \hat{V}^{l} \hat{v}^{l} + \im V^{l} v^{l} + \im Q^{l} q^{l}) \nonumber \\
    \fl \qquad  - \im Q^{L} q^{L} - \sum_{l=1}^{L} \alpha^{l} \log \int \diff \hat{h}^{l} \diff h^{l} \mathrm{Tr}_{\hat{s}^{l}, s^{l}} \mathcal{M}^{l}(\hat{s}^{l}, s^{l}, \hat{h}^{l}, h^{l}), \\
    \fl \mathcal{M}^{l}(\hat{s}^{l}, s^{l}, \hat{h}^{l}, h^{l}) := \frac{ e^{ -\frac{1}{2} (H^{l})^{\transpose} \Sigma_{l}^{-1} H^{l} } }{\sqrt{(2\pi)^2 |\Sigma_{l}}|} P(\hat{s}^{l}|\hat{h}^{l}) P(s^{l}|h^{l}) e^{-\im \hat{V}^{l} (\hat{s}^{l})^2 - \im V^{l} (v^{l})^2 - \im Q^{l} \hat{s}^{l} s^{l} }, \; 1 \leq l < L, \\
    \fl  \mathcal{M}^{L}(\hat{s}^{L}, s^{L}, \hat{h}^{L}, h^{L}) := \frac{ e^{ -\frac{1}{2} (H^{L})^{\transpose} \Sigma_{L}^{-1} H^{L} } }{\sqrt{(2\pi)^2 |\Sigma_{L}}|} \frac{ e^{\beta \hat{s}^{L} \hat{h}^{L} }}{ 2\cosh(\beta \hat{h}^{L}) } \frac{ e^{\beta s^{L} h^{L}} }{ 2\cosh(\beta h^{L}) }  e^{-\im Q^{L} \hat{s}^{L} s^{L} },
\end{eqnarray}
where $\alpha^{0}=1$ since $N^{0}=N$.

Setting the derivatives with respect to the conjugate order parameters $\partial \Phi / \partial \im \hat{V}^{l}$, $\partial \Phi / \partial \im V^{l}$, $\partial \Phi / \partial \im Q^{l}$ to zero yields the  saddle point equations
\begin{eqnarray}
    \fl \hat{v}^{0} = v^{0} = 1, \quad q^{0} = 1, \label{eq:saddle_q0_general} \\
    \fl \hat{v}^{l} = \frac{\int \diff \hat{h}^{l} \diff h^{l} \mathrm{Tr}_{\hat{s}^{l}, s^{l}} \big( \hat{s}^{l} \big)^2 \mathcal{M}^{l}(\hat{s}^{l}, s^{l}, \hat{h}^{l}, h^{l})}{\int \diff \hat{h}^{l} \diff h^{l} \mathrm{Tr}_{\hat{s}^{l}, s^{l}} \mathcal{M}^{l}(\hat{s}^{l}, s^{l}, \hat{h}^{l}, h^{l})} = \langle \big( \hat{s}^{l} \big)^2 \rangle_{\mathcal{M}^{l}}, \quad v^{l} = \langle \big( s^{l} \big)^2 \rangle_{\mathcal{M}^{l}}, \quad 1 \leq l < L, \label{eq:saddle_ql_general} \\
    \fl q^{l} = \frac{\int \diff \hat{h}^{l} \diff h^{l} \mathrm{Tr}_{\hat{s}^{l}, s^{l}} \big( \hat{s}^{l} s^{l} \big) \mathcal{M}^{l}(\hat{s}^{l}, s^{l}, \hat{h}^{l}, h^{l})}{\int \diff \hat{h}^{l} \diff h^{l} \mathrm{Tr}_{\hat{s}^{l}, s^{l}} \mathcal{M}^{l}(\hat{s}^{l}, s^{l}, \hat{h}^{l}, h^{l})} = \langle \hat{s}^{l} s^{l} \rangle_{\mathcal{M}^{l}}, \quad 1 \leq l \leq L, \label{eq:saddle_iQl_general}
\end{eqnarray}
in which $\mathcal{M}^{l}(\hat{s}^{l}, s^{l}, \hat{h}^{l}, h^{l})$ bears the meaning of an effective measure~\cite{Coolen2001}. Notice that $q^{L}$ is an \emph{input parameter} imposing a nonlinear end point constraint on $\im Q^{L}$, which differs from the generating functional analysis calculation of typical behaviors~\cite{BoLi2018}, where $q^{L}$ is a dynamical variable and $\im Q^{L} = 0$ at the saddle point. 

Setting $\partial \Phi / \partial q^{l}$ to zero yields the saddle point equations for the conjugate order parameters $\im Q^{l}$
\begin{eqnarray}
    \im Q^{l-1} = \frac{\alpha^{l}}{\alpha^{l-1}} \frac{\int \diff \hat{h}^{l} \diff h^{l} \mathrm{Tr}_{\hat{s}^{l}, s^{l}} \frac{\partial}{\partial q^{l-1}} \mathcal{M}^{l}(\hat{s}^{l}, s^{l}, \hat{h}^{l}, h^{l})}{\int \diff \hat{h}^{l} \diff h^{l} \mathrm{Tr}_{\hat{s}^{l}, s^{l}} \mathcal{M}^{l}(\hat{s}^{l}, s^{l}, \hat{h}^{l}, h^{l})}, \quad 1 \leq l \leq L. \label{eq:saddle_conjugate_order_parameter}
\end{eqnarray}
Similar relations holds for $\im \hat{V}^{l}$ and $\im V^{l}$. While the conjugate order parameters $\{ \hat{V}^{l}, V^{l}, Q^{l} \}$ are defined on the real axis, they can be extended to the complex plane and evaluated on the imaginary axis in the saddle point approximation, in which case $\{ \im \hat{V}^{l}, \im V^{l}, \im Q^{l} \}$ are real variables.
Other observables can be computed by resorting to the effective measure $\mathcal{M}^{l}$ once the saddle point is obtained, e.g., the mean activations are given by~\cite{Coolen2001}
\begin{equation}
    \hat{m}^{l} = \langle \hat{s}^{l} \rangle_{\mathcal{M}^{l}}, \quad m^{l} = \langle s^{l} \rangle_{\mathcal{M}^{l}}. \label{eq:mean_activation}    
\end{equation}

Since the covariance matrix $\Sigma_{l}(q^{l-1},...)$ depends on the order parameters of layer $l-1$, the effective measure $\mathcal{M}^{l}$ at layer $l$ depends on the order parameters $\{ q^{l-1},... \}$ of the previous layer, while it depends on the conjugate order parameters $\{ \im Q^{l},... \}$ of the current layer. We then observe that the order parameters $\{ q^{l},... \}$ propagate forward in layers, while $\{ \im Q^{l},... \}$ encoding the randomness leading to the desired deviation propagate backward, which resembles the structure in optimal control problem~\cite{Grafke2019}. Therefore, we solve the saddle point equations in a forward-backward iteration manner until convergence. Another feature to notice in Eq.~(\ref{eq:saddle_conjugate_order_parameter}) is the dependence of the saddle point solution on the layer-shape parameters $\{ \alpha^{l} \}$, which does not play a role in the mean field solutions where all the conjugate order parameters $\{ \im Q^{l},... \}$ vanish~\cite{BoLi2018}.

\subsection{Explicit solutions for sign and ReLU activation functions}
For networks with sign activation function the order parameters satisfy $\hat{v}^{l}=v^{l}=1$, such that the only meaningful order parameters are $\{ q^{l}, Q^{l} \}$. The potential function $\Phi$ can be computed analytically, taking the form 
\begin{eqnarray}
    \fl \Phi(\boldsymbol{Q}, \boldsymbol{q}|q^{L}) &= - \alpha^{0} \im Q^{0} (q^{0} - 1) -\sum_{l=1}^{L} \alpha^{l} \im Q^{l} q^{l} \nonumber \\
    \fl & \quad -\sum_{l=1}^{L} \alpha^{l} \log \bigg[ \cosh(\im Q^{l}) - \sinh(\im Q^{l}) \frac{2}{\pi} \sin^{-1}(\sqrt{1-(\eta^{l})^2} q^{l-1}) \bigg],
\end{eqnarray}
while the saddle point equations become
\begin{eqnarray}
    q^{0} = 1, \\
    q^{l} = \frac{ -\sinh(\im Q^{l}) + \cosh(\im Q^{l}) \frac{2}{\pi} \sin^{-1}(\sqrt{1-(\eta^{l})^2} q^{l-1}) }{ \cosh(\im Q^{l}) - \sinh(\im Q^{l}) \frac{2}{\pi} \sin^{-1}(\sqrt{1-(\eta^{l})^2} q^{l-1}) }, \quad \forall 1 \leq l \leq L, \label{eq:saddle_ql_sign} \\
    \im Q^{l-1} = \frac{ \frac{2}{\pi} \sinh(\im Q^{l}) }{ \cosh(\im Q^{l}) - \sinh(\im Q^{l})\frac{2}{\pi}\sin^{-1}(\sqrt{1-(\eta^{l})^2} q^{l-1}) } \nonumber \\
    \qquad \qquad \times \frac{ \alpha^{l} \sqrt{1-(\eta^{l})^2} }{ \alpha^{l-1} \sqrt{1-[1-(\eta^{l})^2](q^{l-1})^2}  }, \quad \forall 1 \leq l \leq L.
\end{eqnarray}
Note that $q^{L}$ in Eq.~(\ref{eq:saddle_ql_sign}) is an input parameter.

For networks with ReLU activation function the potential function $\Phi$ also admits an explicit expression 
\begin{eqnarray}
    \fl \Phi(\boldsymbol{Q},\boldsymbol{q},\hat{\boldsymbol{V}},\hat{\boldsymbol{v}},\boldsymbol{V},\boldsymbol{v}|q^{L}) = - \alpha^{0} \big[ \im \hat{V}^{0} (\hat{v}^{0}-1) + \im V^{0} (v^{0}-1) +  \im Q^{0} (q^{0}-1) \big] \nonumber \\
    \fl -\sum_{l=1}^{L-1} \alpha^{l} (\im \hat{V}^{l} \hat{v}^{l} + \im V^{l} v^{l} + \im Q^{l} q^{l}) - \im Q^{L} q^{L} \nonumber \\
    \fl \!-\!\sum_{l=1}^{L-1} \alpha^{l} \log \left\{ \frac{1}{2\pi\sqrt{|\Sigma_{l}}|} \left[ \frac{1}{\sqrt{|A^{l}|}} \left( \frac{\pi}{2} \!-\! \tan^{-1}\!\left( \frac{A^{l}_{12}}{\sqrt{|A^{l}|}} \right) \right) \!+\!\frac{1}{\sqrt{|B^{l}|}} \left( \frac{\pi}{2} \!+\! \tan^{-1}\!\left( \frac{B^{l}_{12}}{\sqrt{|B^{l}|}} \right) \right) \right. \right. \nonumber \\
    \fl \qquad \left. \left. + \frac{1}{\sqrt{|\Sigma^{-1}_{l}|}} \left( \frac{\pi}{2} - \tan^{-1}\left( \frac{\Sigma^{-1}_{l,12}}{\sqrt{|\Sigma^{-1}_{l}|}} \right) \right) +\frac{1}{\sqrt{|C^{l}|}} \left( \frac{\pi}{2} + \tan^{-1}\left( \frac{C^{l}_{12}}{\sqrt{|C^{l}|}} \right) \right) \right] \right\} \nonumber \\
    \fl - \alpha^{L} \log \left[ \cosh(\im Q^{L}) - \sinh(\im Q^{L}) \frac{2}{\pi} \tan^{-1}\left( \frac{\Sigma_{L,12}}{\sqrt{|\Sigma_{L}|}} \right) \right],
\end{eqnarray}
where $A^{l},B^{l},C^{l}$ are $2\times 2$ matrices defined as
\begin{equation}
    \fl A^{l} = \Sigma^{-1}_{l} + \left[ \matrix{2\im \hat{V}^{l} & \im Q^{l} \cr \im Q^{l} & 2\im V^{l}} \right], \quad B^{l} = \Sigma^{-1}_{l} + \left[ \matrix{0 & 0 \cr 0 & 2\im V^{l}} \right], \quad C^{l} = \Sigma^{-1}_{l} + \left[ \matrix{2\im \hat{V}^{l} & 0 \cr 0 & 0} \right].
\end{equation}
The saddle point equations also admit a close-form expression accordingly.

\section{Large deviations in input sensitivity of functions} \label{sec:LDP_input}
In probing the sensitivity of a function to the flipping of input variables, the weights of two networks considered are taking the same values $\boldsymbol{w} = \boldsymbol{\hat{w}}$, which is done by setting $\eta^{l}=0$ in Eq.~(\ref{eq:def_eta_perturbation}). We constrain the input $\boldsymbol{s}^{0}$ of the perturbed system to have a pre-defined overlap $q^{0}$ (or Hamming distance $N^{0}(1-q^{0})/2$) with the input $\boldsymbol{\hat{s}}^{0}$ of the reference system. The sensitivity of the output overlaps to input perturbations is investigated through the conditional probability
\begin{equation}
    P(q^{L}|q^{0}) = \frac{P(q^{L}, q^{0})}{P(q^{0})} = \frac{\left\langle \delta\bigg( \frac{1}{N^{L}}\sum_{i} \hat{s}^{L}_{i} s^{L}_{i} - q^{L} \bigg) \delta\bigg( \frac{1}{N^{0}}\sum_{i} \hat{s}^{0}_{i} s^{0}_{i} - q^{0} \bigg) \right\rangle}{\left\langle \delta\bigg( \frac{1}{N^{0}}\sum_{i} \hat{s}^{0}_{i} s^{0}_{i} - q^{0} \bigg) \right\rangle}. \label{eq:PqL_given_q0}
\end{equation}

Without loss of generality, we choose a decoupled input distribution $P(\boldsymbol{\hat{s}}^{0}, \boldsymbol{s}^{0}) = \prod_{i} P(\hat{s}^{0}_{i}) P(s^{0}_{i}) = \prod_{i} (\frac{1}{2}\delta_{\hat{s}_{i}^{0},1}+\frac{1}{2}\delta_{\hat{s}_{i}^{0},-1}) (\frac{1}{2}\delta_{s_{i}^{0},1}+\frac{1}{2}\delta_{s_{i}^{0},-1})$ while the delta function involving $q^{0}$ in Eq.~(\ref{eq:PqL_given_q0}) constrains the systems to have the desired input correlation. The probability of input overlap $P(q^{0})$ can be computed as
\begin{eqnarray}
    P(q^{0}) = \mathrm{Tr}_{\hat{\boldsymbol{s}}^{0}, \boldsymbol{s}^{0}} \prod_{i} P(\hat{s}^{0}_{i}) P(s^{0}_{i}) \int \frac{\diff Q^{0}}{2\pi/N^{0}} e^{ \im N^{0} Q^{0} \big( q^{0} - \frac{1}{N^{0}} \sum_{i} \hat{s}^{0}_{i} s^{0}_{i} \big) } \nonumber \\
    \qquad\; = \int \frac{\diff Q^{0}}{2\pi/N^{0}} \exp \bigg[ N^{0} \big( \im Q^{0} q^{0} + \log\cosh(\im Q^{0}) \big) \bigg] \nonumber \\
    \qquad\; \approx \exp \bigg[ N^{0} \big( \im Q^{0*} q^{0} + \log\cosh(\im Q^{0*}) \big) \bigg] \nonumber \\
    \qquad\; =: \exp\big[ - N\Phi_{\mathrm{P}} (\im Q^{0*} | q^{0}) \big], \\
    \Phi_{\mathrm{P}} (\im Q^{0} | q^{0}) := -\alpha^{0} \big( \im Q^{0} q^{0} + \log\cosh(\im Q^{0}) \big), \label{eq:potential_of_Pq0} \\
    \im Q^{0*}  := -\tanh^{-1}(q^{0}), \label{eq:saddle_iQ0_of_Pq0}
\end{eqnarray}
where we have made use of the saddle point approximation of $P(q^{0})$ in the large $N^{0}$ limit, with the corresponding potential function defined in Eq.~(\ref{eq:potential_of_Pq0}) and the saddle point solution $\im Q^{0*}$ given in Eq.~(\ref{eq:saddle_iQ0_of_Pq0}).

The computation of the joint probability $P(q^{L}, q^{0})$ is analogous to that of $P(q^{L})$ in earlier sections,
\begin{eqnarray}
    \fl P(q^{L}, q^{0}) = \mathbb{E}_{\hat{\boldsymbol{w}}, \boldsymbol{w}} \mathrm{Tr}_{\boldsymbol{\hat{s}}, \boldsymbol{s}} P(\hat{\boldsymbol{s}}^{0}) \prod_{i=1}^{N^{0}}\delta_{s_{i}^{0},\hat{s}_{i}^{0}}\prod_{l=1}^{L} P(\hat{\boldsymbol{s}}^{l}|\hat{\boldsymbol{w}}^{l},\hat{\boldsymbol{s}}^{l-1}) P(\boldsymbol{s}^{l}|\boldsymbol{w}^{l},\boldsymbol{s}^{l-1}) \nonumber \\
    \fl \qquad\qquad\quad \times \int \frac{\diff Q^{0}}{2\pi/N^{0}} \frac{\diff Q^{L}}{2\pi/N^{L}} e^{ \im N^{0} Q^{0} \big( q^{0} - \frac{1}{N^{0}} \sum_{i} \hat{s}^{0}_{i} s^{0}_{i} \big) + \im N^{L} Q^{L} \big( q^{L} - \frac{1}{N^{L}} \sum_{i} \hat{s}^{L}_{i} s^{L}_{i} \big) } \nonumber \\
    \fl \qquad\qquad = \int \{ \diff \boldsymbol{Q} \diff \boldsymbol{q} ... \} \exp[-N\Phi_{\mathrm{J}}(\boldsymbol{Q},\boldsymbol{q},...|q^{L},q^{0})], \\
    \fl \Phi_{\mathrm{J}} = - \alpha^{0} \big[ \im \hat{V}^{0} (\hat{v}^{0}-1) + \im V^{0} (v^{0}-1) + \big( \im Q^{0} q^{0} + \log\cosh(\im Q^{0}) \big) \big] - \im Q^{L} q^{L} \nonumber \\ 
    \fl \qquad -\sum_{l=1}^{L-1} \alpha^{l} (\im \hat{V}^{l} \hat{v}^{l} + \im V^{l} v^{l} + \im Q^{l} q^{l})  - \sum_{l=1}^{L} \alpha^{l} \log \int \diff \hat{h}^{l} \diff h^{l} \mathrm{Tr}_{\hat{s}^{l}, s^{l}} \mathcal{M}^{l}(\hat{s}^{l}, s^{l}, \hat{h}^{l}, h^{l}).
\end{eqnarray}
The saddle point of $\im Q^{0}$ satisfies $\im Q^{0*} = -\tanh^{-1}(q^{0})$, which coincides with the one of $P(q^{0})$ in Eq.~(\ref{eq:saddle_iQ0_of_Pq0}). So the conditional distribution satisfies
\begin{eqnarray}
    \fl P(q^{L}|q^{0}) \approx \exp\big[ -N\Phi(\boldsymbol{Q}^{*}, \boldsymbol{q}^{*},...|q^{L}, q^{0}) \big] = \exp\big[ -N(\Phi_{\mathrm{J}}^{*} - \Phi_{\mathrm{P}}^{*}) \big] \nonumber \\
    \fl \Phi(\boldsymbol{Q}, \boldsymbol{q},...|q^{L}, q^{0}) = - \alpha^{0} \big[ \im \hat{V}^{0} (\hat{v}^{0}-1) + \im V^{0} (v^{0}-1) \big] - \im Q^{L} q^{L} \nonumber \\ 
    \fl \qquad -\sum_{l=1}^{L-1} \alpha^{l} (\im \hat{V}^{l} \hat{v}^{l} + \im V^{l} v^{l} + \im Q^{l} q^{l})  - \sum_{l=1}^{L} \alpha^{l} \log \int \diff \hat{h}^{l} \diff h^{l} \mathrm{Tr}_{\hat{s}^{l}, s^{l}} \mathcal{M}^{l}(\hat{s}^{l}, s^{l}, \hat{h}^{l}, h^{l}),
\end{eqnarray}
where the saddle point solution $\{ \boldsymbol{Q}^{*}, \boldsymbol{q}^{*},... \}$ have the same form as those in Sec.~\ref{sec:saddle_point_eq}, except that $q^{0} = 1$ in Eq.~(\ref{eq:saddle_q0_general}) is replaced by the pre-defined value $q^{0}$ under investigation.  

\section{Results} \label{sec:result}
\subsection{Weight sparsification}
\begin{figure}
    \centering
    \subfloat[]{\includegraphics[scale=0.37]{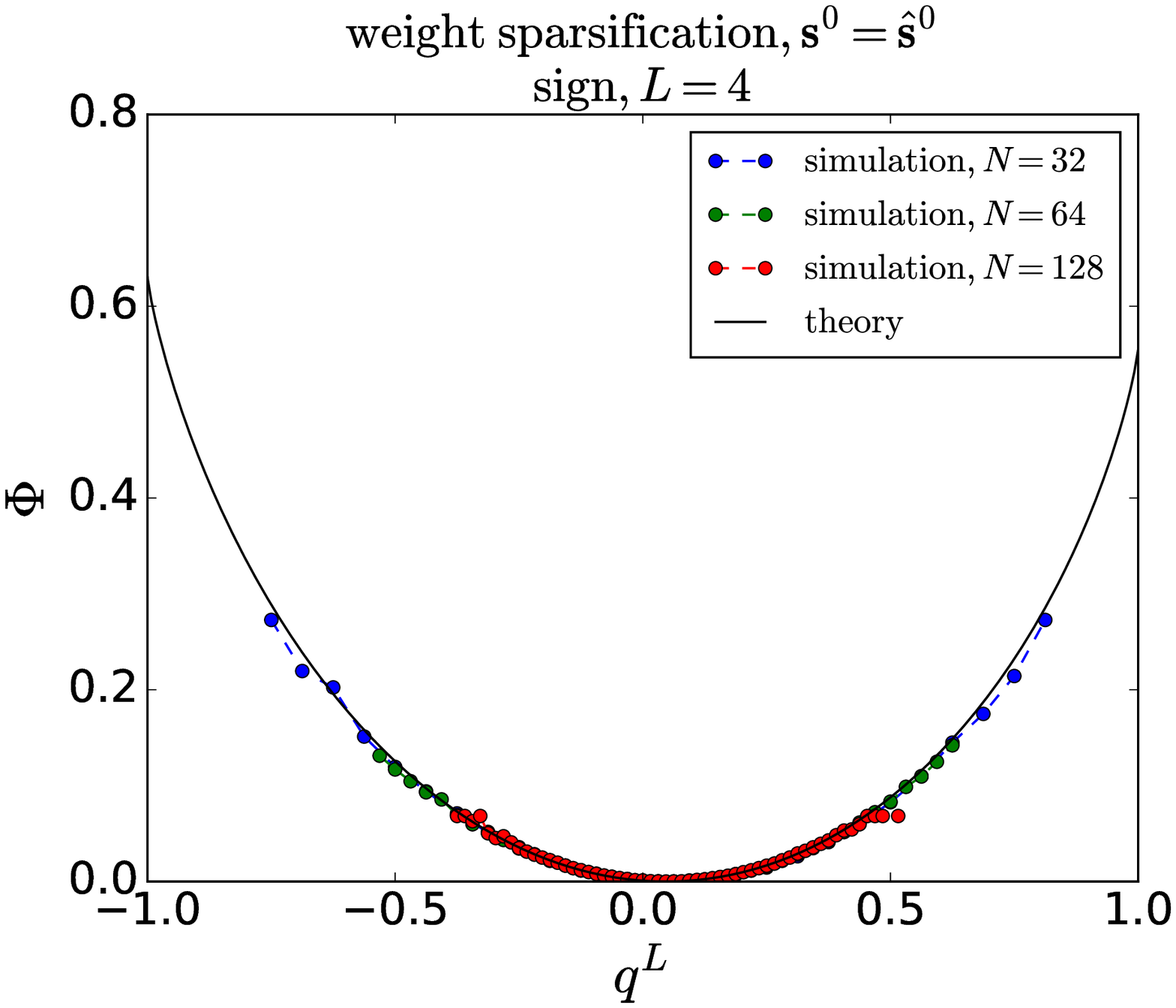}}
    \subfloat[]{\includegraphics[scale=0.37]{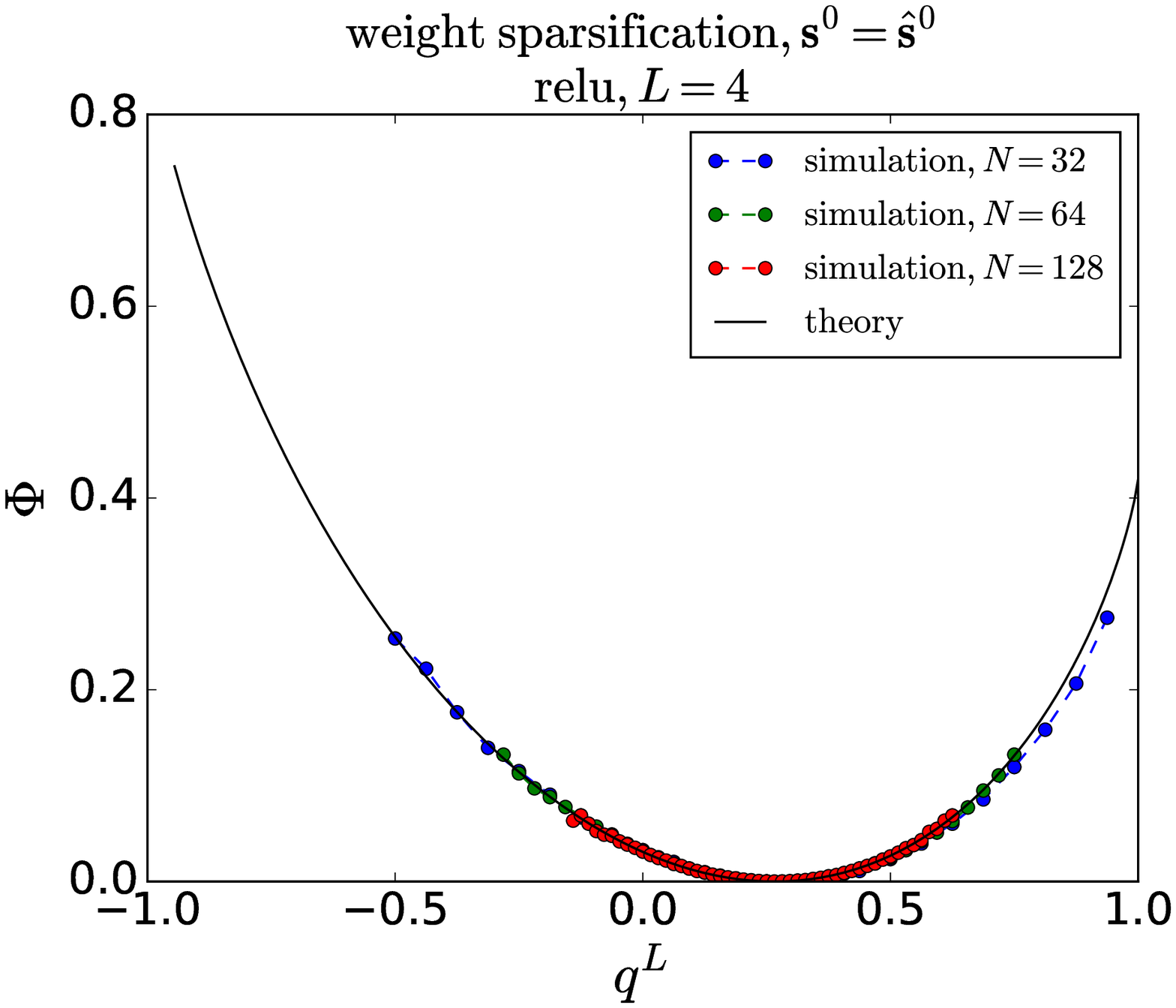}} \\
    \subfloat[]{\includegraphics[scale=0.37]{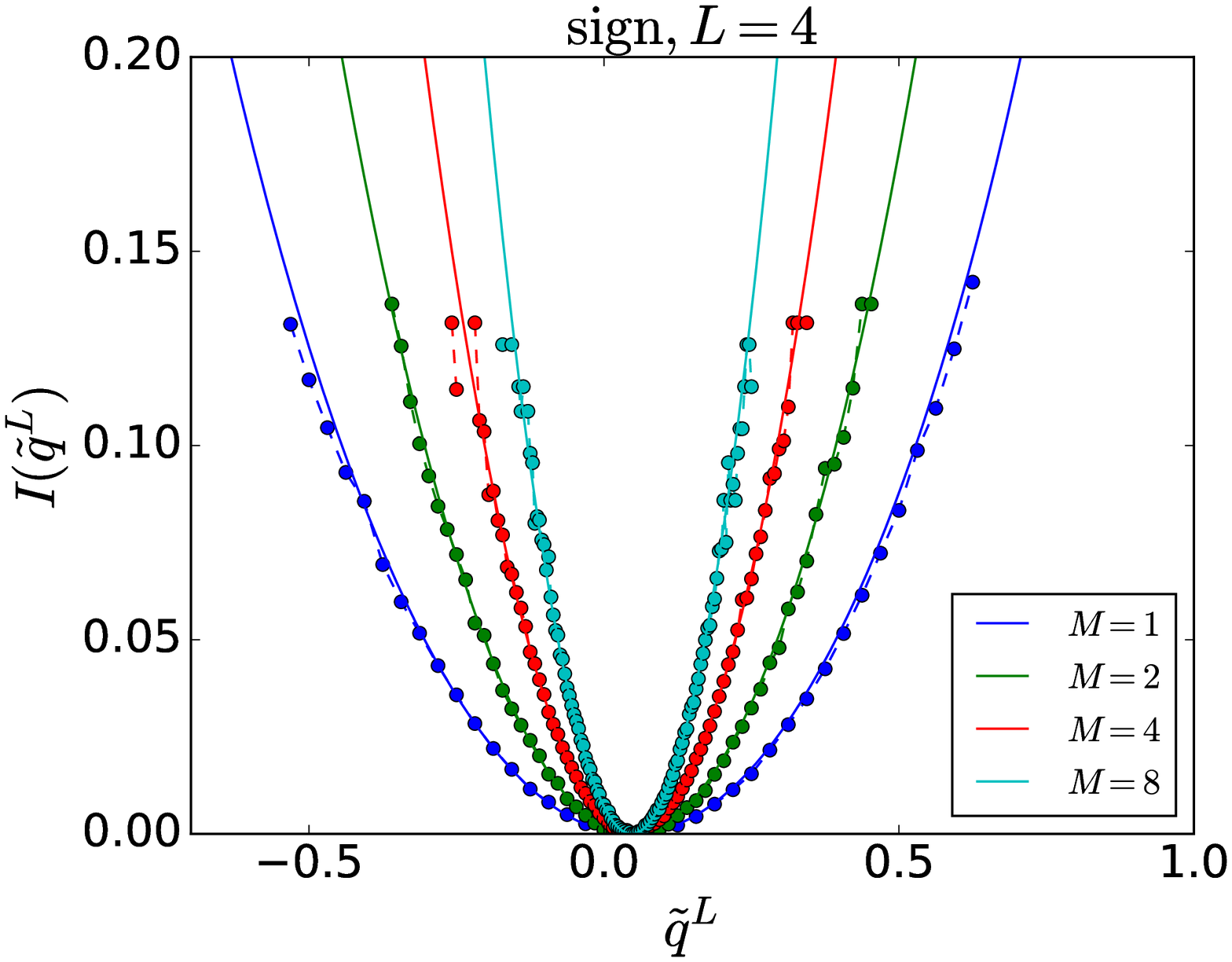}}
    \subfloat[]{\includegraphics[scale=0.37]{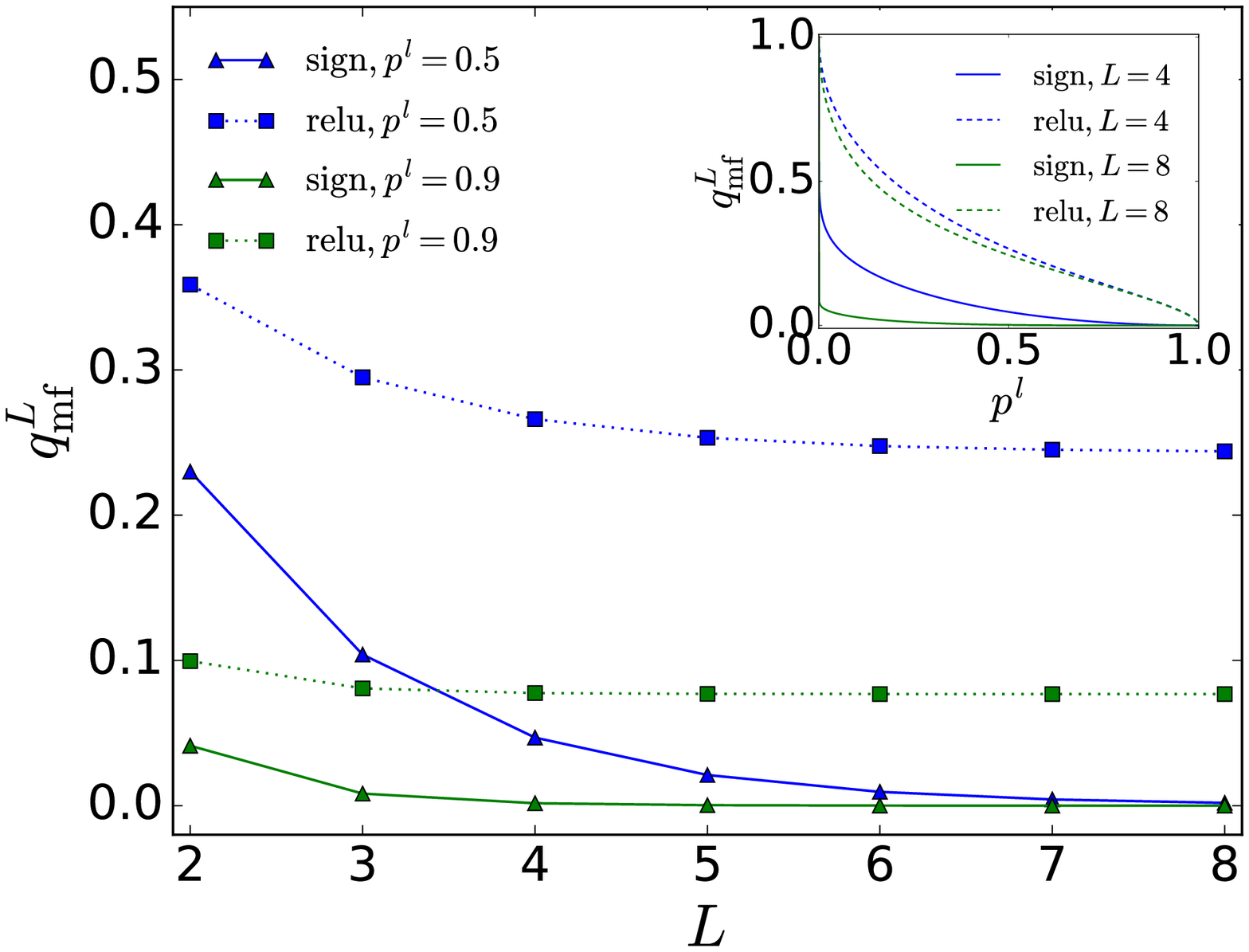}}
    \caption{Weight sparsification of random DNN. In (a)(b)(c), we set $L=4$ and $p^{l}=1/2$; solid lines correspond to theory while dashed lines with circle markers correspond to estimation from simulation. The estimation of the rate function from simulations are obtained by $100,000$ samples and the corresponding curve has been shifted such that the minimum is at zero. (a) The rate function $\Phi$ vs $q^{L}$ for sign activation function. (b) The rate function $\Phi$ vs $q^{L}$ for ReLU activation function. (c) The rate function $I(\tilde{q}^{L})$ of output overlap $\tilde{q}^{L}$ defined by $M$ patterns; the theoretical results are given by Eq.~(\ref{eq:rate_func_tilde_qL}), while the simulation results are obtained on systems with $N=64$. (d) Mean field solutions of output overlap $q^{L}_{\mathrm{mf}}$ as a function of system depth $L$. Inset: $q^{L}_{\mathrm{mf}}$ vs $p^{l}$ for different depths. }
    \label{fig:sparse_weight}
\end{figure}

We first consider the effect of weight perturbation by sparsifying connections as in Eq.~(\ref{eq:def_sparse_perturbation}). For a concrete example, we consider DNN with $L=4$, uniform layer width $\alpha^{l}=1$ and disconnection probability $p^{l}=1/2$, for which we compute the large deviation rate function $I(q^{L}) = \Phi(\boldsymbol{Q}^{*}, \boldsymbol{q}^{*}, ...|q^{L})$ by solving the saddle point equation in Sec.~\ref{sec:saddle_point_eq} and compare it to numerical experiments. For relu-DNN, we always set $\sigma_{w}=\sqrt{2}$. The results are shown in Fig.~\ref{fig:sparse_weight}(a)(b), which exhibit a perfect match between the theory and simulation. The most probable $q^{L}$, located at the minimum of $\Phi$ corresponds to the mean field solution, where $q^{L}_{\mathrm{mf}} \approx 0.047$ for sign-DNN and $q^{L}_{\mathrm{mf}} \approx 0.266$ for the relu-DNN. However, in finite systems they have a non-zero probability of admitting a higher value of $q^{L}$ due to fluctuations. We can compute the probability from the rate function by $P(q^{L})=\exp(-N \Phi^{*}(q^{L}))/Z$\footnote{For finite $N^{L}$, the output overlap is a discrete variable $q^{L}\in \{1, 1-\frac{2}{N^{L}}, 1-\frac{4}{N^{L}}, ..., -1\}$, so it is convenient to consider the discretized probability distribution of $q^{L}$ as $\mathrm{Prob}(q^{L}) = P(q^{L}) \Delta q^{L}=\exp(-N \Phi^{*}(q^{L}))/Z$; the normalization constant is computed as $Z = \sum_{k} \exp(-N \Phi^{*}(q^{L}_{k})) \Delta q^{L}$, where the summation runs over all possible values of $q^{L}$ and $\Delta q^{L} = \frac{2}{N^{L}}$. Although we could not find the saddle point solution of $\Phi(...|q^{L})$ in the vicinity of $q^{L}=-1$ for relu-DNN (see Fig.~\ref{fig:sparse_weight}(b)), the contribution from that region to the cumulative probability of the overlap is negligible .} and estimate the tail probability of output mismatch. As an example we consider $N=64$ and find that $P(q^{L}>1/2) \approx 0.055\%$ for sign-DNN and $P(q^{L}>1/2) \approx 3.8\%$ for relu-DNN, which is non-negligible especially for ReLU activation.\footnote{Notice that such estimation is obtained by saddle point approximation in Eq.~(\ref{eq:PqL_eq_intexpNPhi}) and by keeping the leading order contribution, which may be slightly biased for small $N$.}

In Fig.~\ref{fig:sparse_weight}(c), we also demonstrate that the approximation of rate function $I(\tilde{q}^{L})$ of output overlap $\tilde{q}^{L}$, estimated for $M$ patterns by employing Eq.~(\ref{eq:rate_func_tilde_qL}), is accurate for DNN with sign activation, while the approximation does not hold for deep ReLU networks (see~\ref{sec:appendix_multiple_input}). Therefore in sign-DNN, the probability of finding perturbed DNN agreeing on all $M$ patterns with the reference DNN decays exponentially with $M$ (at least for small $M$ values). This may not be the case in relu-DNN which requires further exploration in a future study.

In Fig.~\ref{fig:sparse_weight}(d), we compare the mean field output overlaps $q^{L}_{\mathrm{mf}}$ between DNN with sign and ReLU activations for different system depths and disconnection probability $p^{l}$. It is shown that relu-DNN are more robust to weight sparsification perturbation, as expected; the perturbed relu-DNN have residual correlations with the reference networks even after removing $90\%$ of the weights. The robustness of relu-DNN to weight dilution was also observed and theoretically analysed in \cite{Haiping2018}. Finally, we remark that our scenario is different from the practical methods used to prune networks trained on specific data; in this case particular heuristic rules have been developed to disconnect weights instead of the random removal used here. The success of weight pruning in practice hightlights the weight-redundancy in real trained networks~\cite{LeCun1989, Haiping2018} but may also be influenced by properties of the data used and training methods. This behaviour is absent in random networks with random data, as indicated in the inset of Fig.~\ref{fig:sparse_weight}(d), where even a small dilution probability can deteriorate the overlap. Additional modelling considerations are needed to address practical scenarios.

\subsection{Weight binarization}
\begin{figure}
    \centering
    \subfloat[]{\includegraphics[scale=0.37]{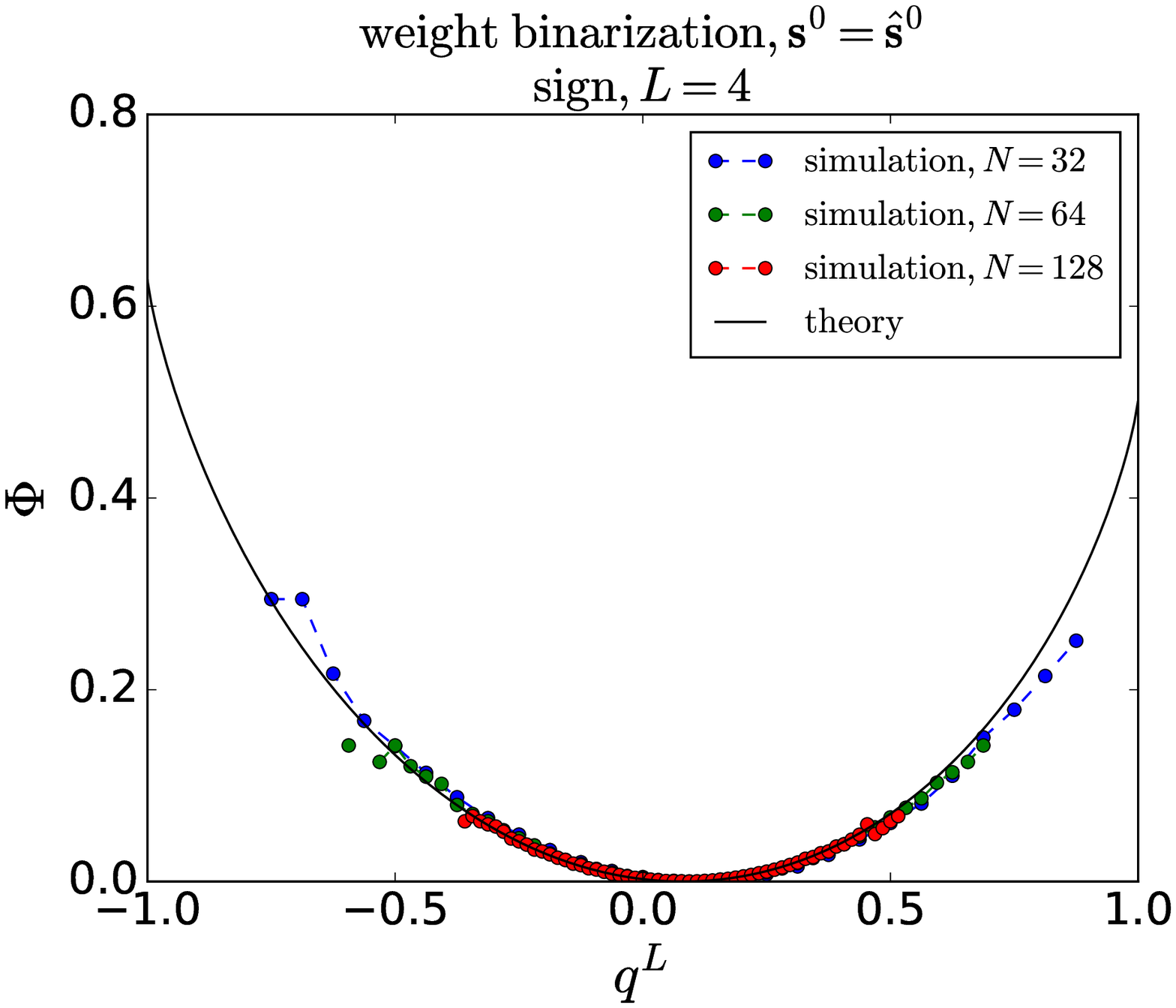}}
    \subfloat[]{\includegraphics[scale=0.37]{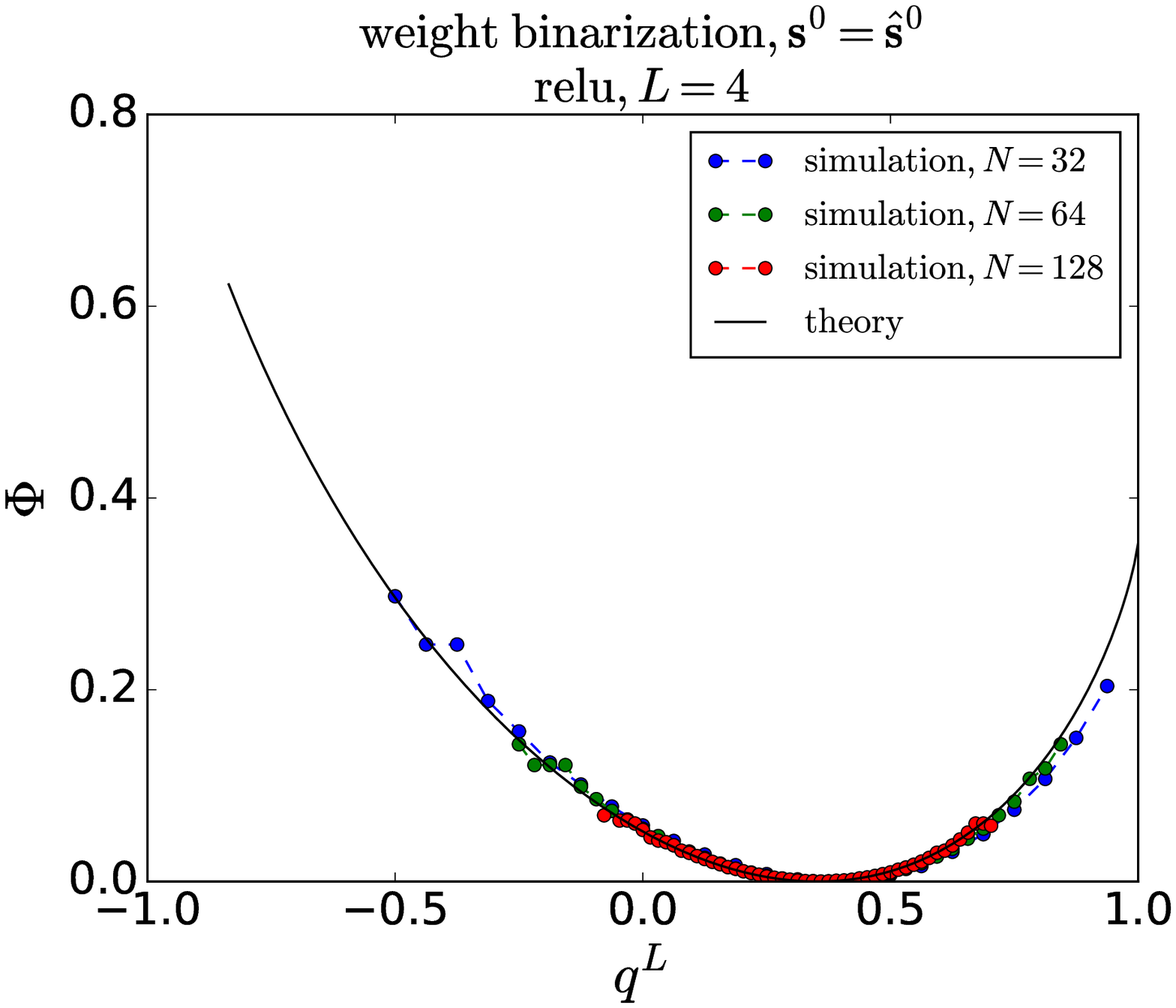}} \\
    \subfloat[]{\includegraphics[scale=0.37]{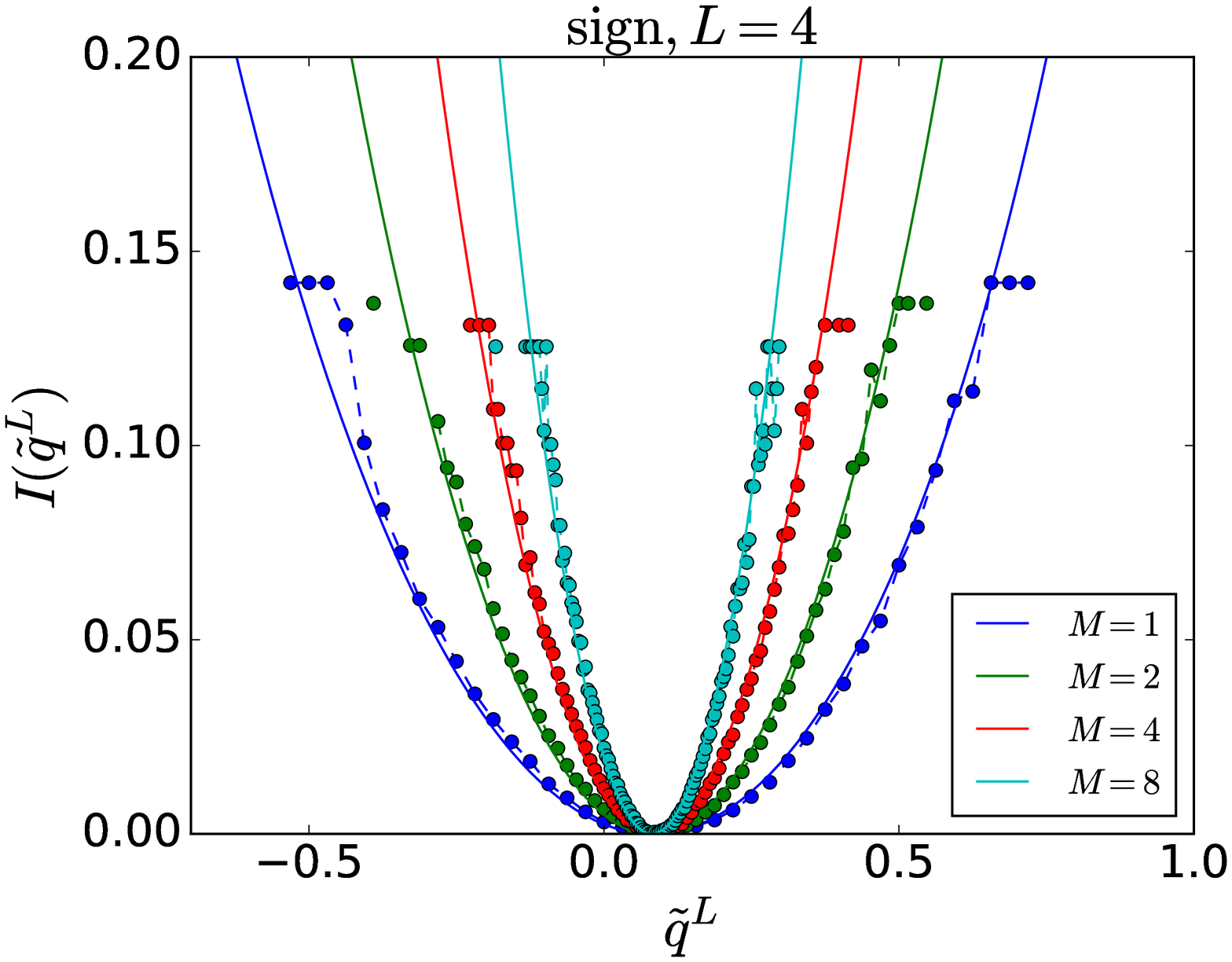}}
    \subfloat[]{\includegraphics[scale=0.37]{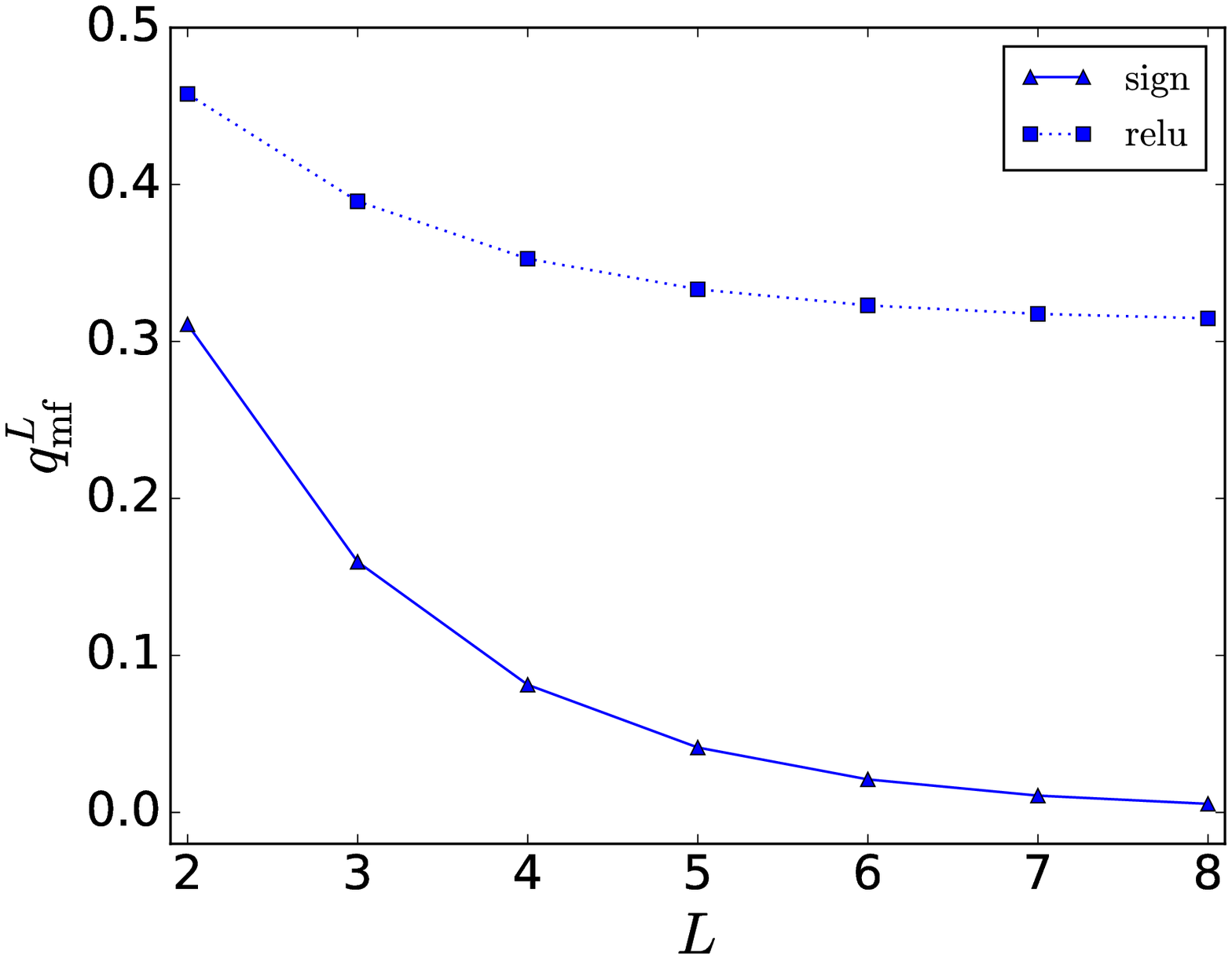}}
    \caption{Weight binarization of random DNN. (a) $\Phi$ vs $q^{L}$ for sign activation function. (b) $\Phi$ vs $q^{L}$ for ReLU activation. (c) The rate function $I(\tilde{q}^{L})$ of output overlap $\tilde{q}^{L}$ defined by $M$ patterns; solid lines are theoretical results while dashed lines with circle markers are estimated by simulation. (d) Mean field solutions of output overlap $q^{L}_{\mathrm{mf}}$ as a function of system depth $L$.}
    \label{fig:binary_weight}
\end{figure}

We then consider the effect of perturbation by binarization of weight variables as in Eq.~(\ref{eq:def_binary_perturbation}). Also here we consider uniform layer width $\alpha^{l}=1$. The results shown in Fig.~\ref{fig:binary_weight}, are very similar to the effect of weight sparsification. As pointed out in Sec.~\ref{sec:unify_perturbation}, binarizing weights of random DNN corresponds to rotating the weight vector $\hat{\boldsymbol{w}}^{l}_{i}$ by an angle $\theta^{l}=\cos^{-1}\sqrt{\frac{2}{\pi}}$~\cite{Anderson2018}, or equivalently, disconnecting weights with a particular probability $p^{l} = 1 - \frac{2}{\pi}$. The matches between theory and simulation in Fig.~\ref{fig:binary_weight}(a)(b)(c) validates the large deviation-based analysis in both sign and relu-DNN and the scaling relation of Eq.~(\ref{eq:rate_func_tilde_qL}) in sign-DNN. The relu-DNN are more biased to the regime of positive correlation and more robust to binarizing perturbation as seen in Fig.~\ref{fig:binary_weight}(d).

\subsection{Sensitivity to input perturbation}
We have shown that relu-DNN with random weights are robust to parameter perturbations such as weight sparsification and weight binarization, which is a desired property for better generalization. On the other hand, such network ensembles typically represent simple functions as studied in~\cite{Valle-Perez2018, Palma2019}. The simplicity of the functions generated is one reason accounting for the observed robustness to parameter perturbation.

\begin{figure}
    \centering
    \subfloat[]{\includegraphics[scale=0.38]{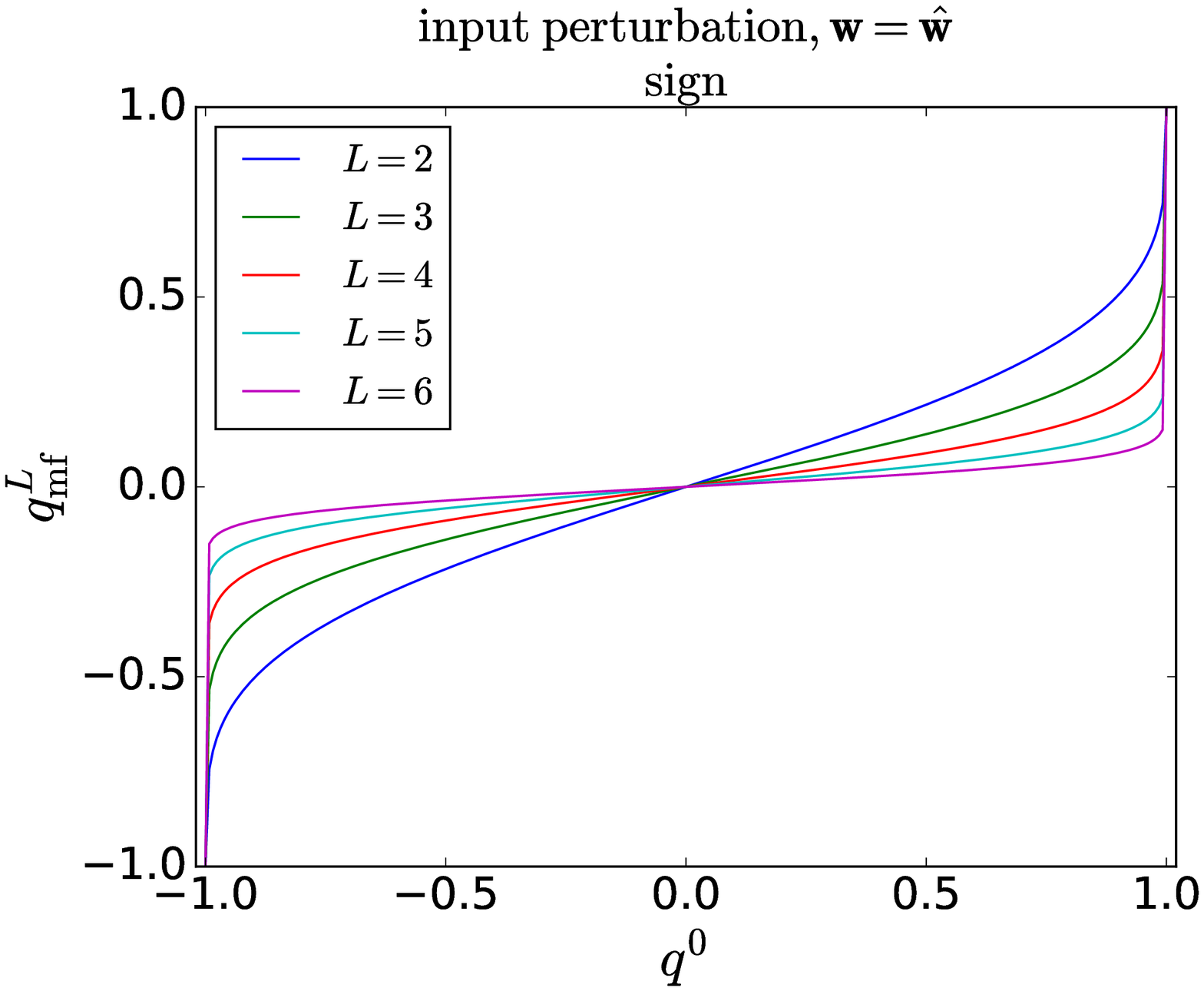}}
    \subfloat[]{\includegraphics[scale=0.38]{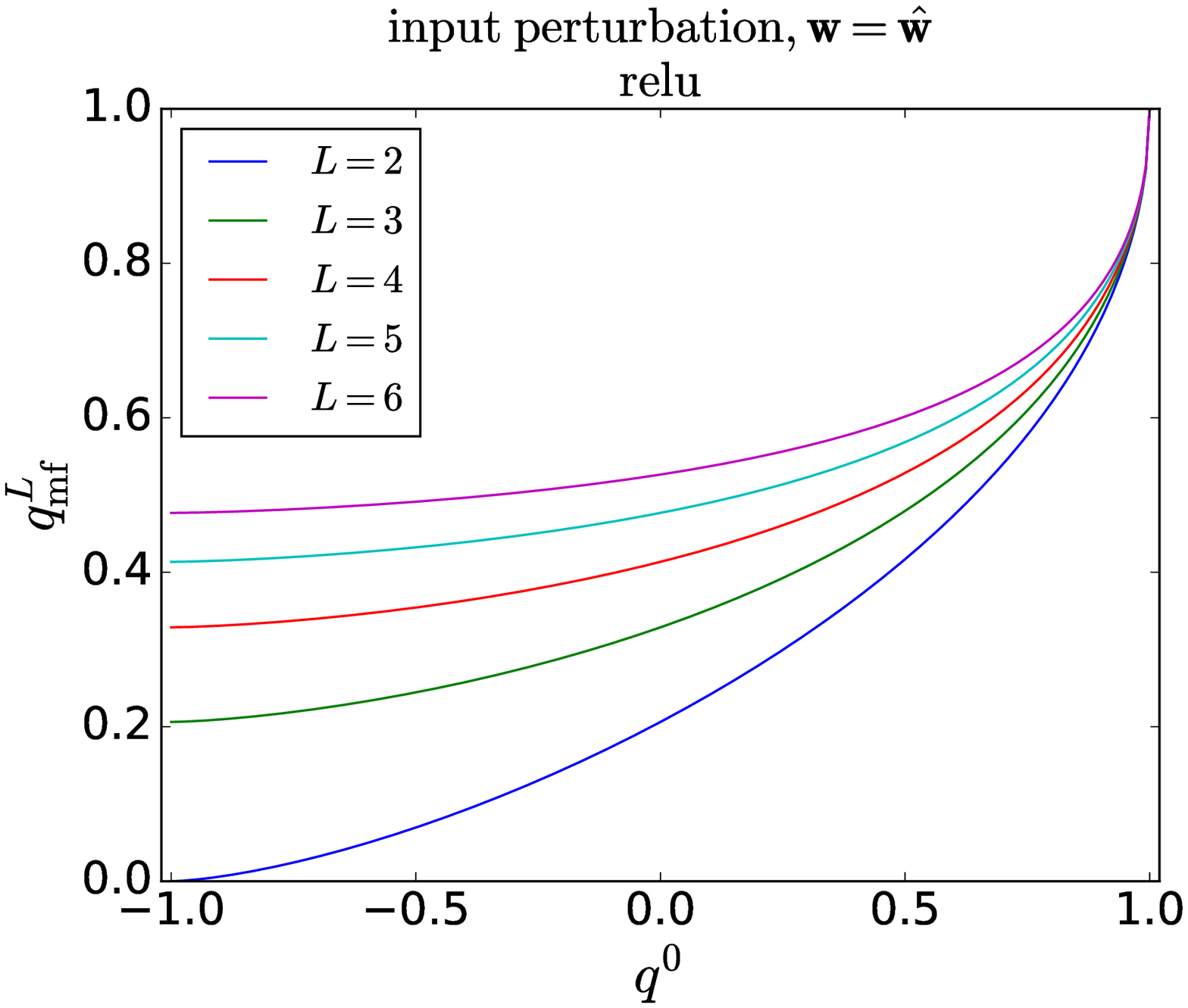}} \\
    \subfloat[]{\includegraphics[scale=0.38]{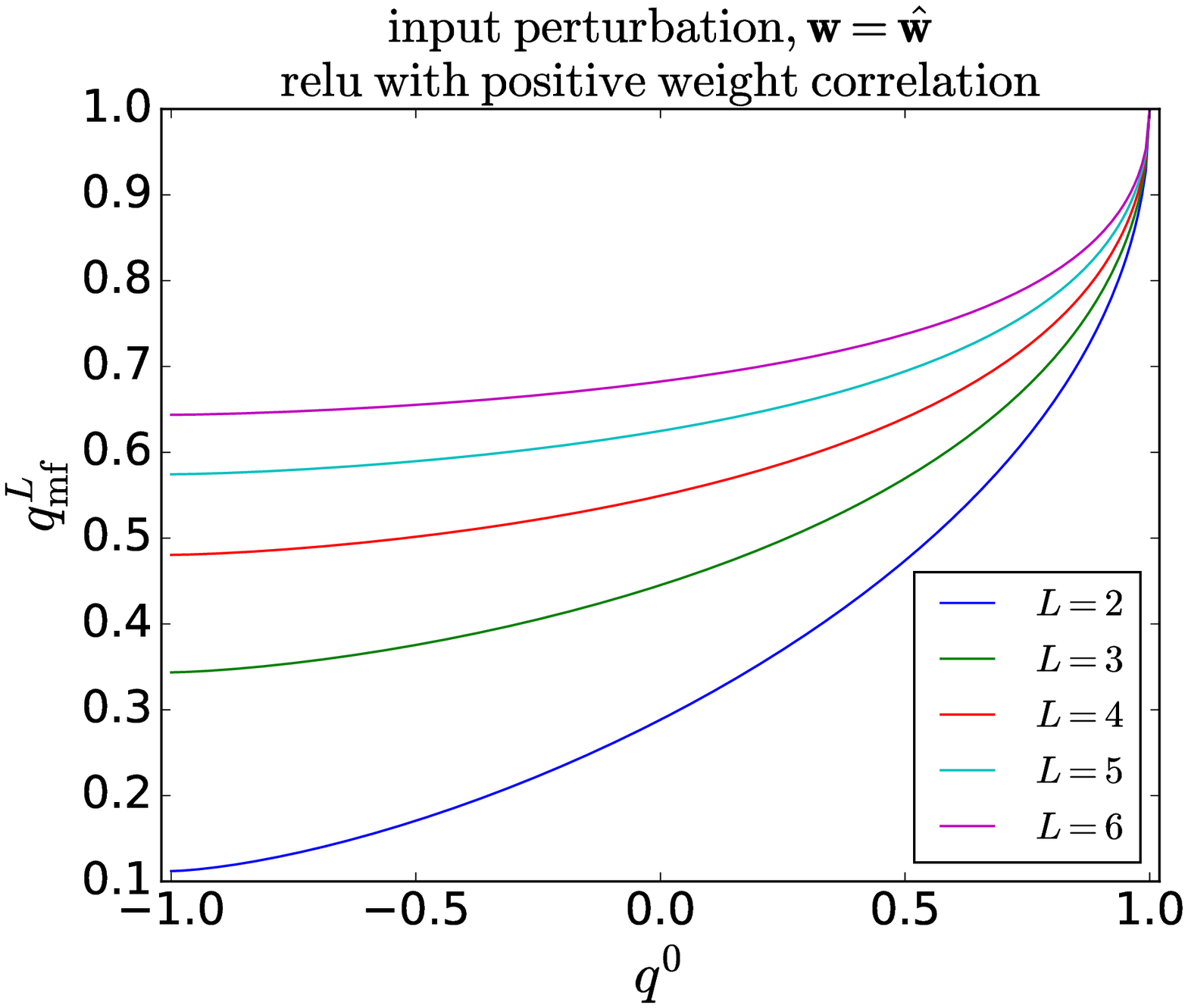}}
    \subfloat[]{\includegraphics[scale=0.38]{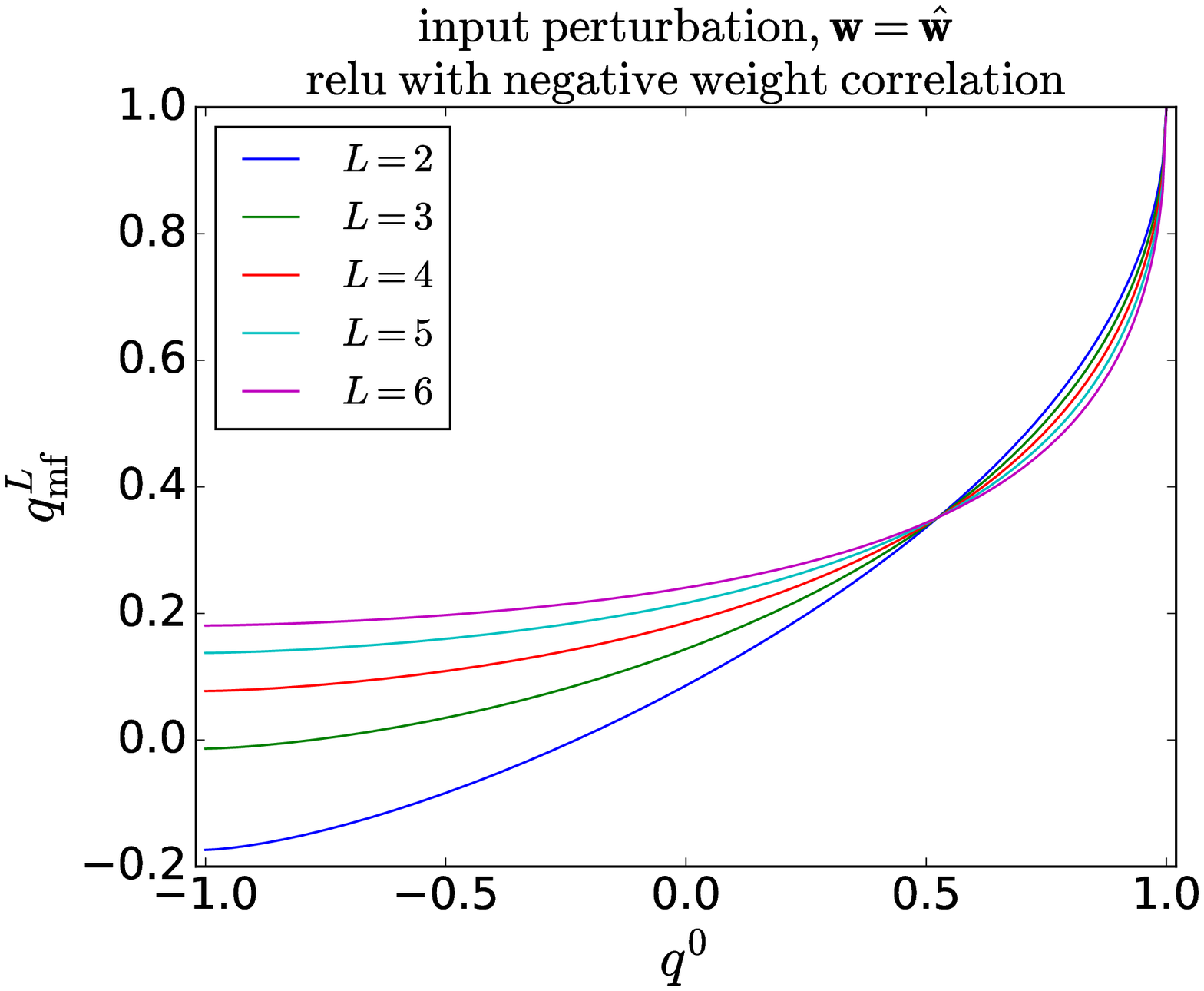}}
    \caption{ Mean field solutions $q^{L}_{\mathrm{mf}}$ vs $q^{0}$ in the scenario of input perturbation where $\boldsymbol{w} = \hat{\boldsymbol{w}}$. In all architectures, sign activation function is applied at the output layer. (a) DNN with sign activation functions and uncorrelated random weights. (b) DNN with ReLU activation at the hidden layers, with  uncorrelated random weights, and sign activation at the output layer. (c) Relu-DNN with positive weight correlation $c=2/(3N)$. (d) Relu-DNN with negative weight correlation $c=-2/(3N)$.} 
    \label{fig:qL_mf_vs_q0}
\end{figure}

To probe the function complexity, we study the function sensitivity under input perturbation while keeping $\boldsymbol{w}=\hat{\boldsymbol{w}}$~\cite{Franco2006}. Flipping $n$ input variables corresponds to the input overlap $q^{0} = 1 - \frac{2n}{N^{0}}$. In Fig.~\ref{fig:qL_mf_vs_q0}(a) and (b) we depict the overlap $q^{L}_{\mathrm{mf}}$ of the final output as a function of input overlap $q^{0}$ (keeping in mind that we always apply the sign activation in the output layer). While the outputs become more de-correlated in deeper layers of sign-DNN, the relu-DNN induce correlation at deeper layers. Therefore, random relu-DNN tend to forget the input structure at deeper layers, generating increasingly simpler functions that are robust to parameter perturbation. This phenomenon has been noticed in the Gaussian process-like analysis of DNN~\cite{Duvenaud2014, Daniely2016, Lee2018}.

In~\cite{BoLi2018}, we investigated the effect of weight correlation in the form of $P(\hat{\boldsymbol{w}}^{l}_{i}) = \exp(-\frac{1}{2} (\hat{\boldsymbol{w}}^{l}_{i})^{\transpose} A^{-1} \hat{\boldsymbol{w}}^{l}_{i} ) / \sqrt{(2\pi)^{N^{l-1}}|A|}$, with $A=\sigma^{2}_{w}(I - c J)$ where $I$ is the identity matrix and $J$ the all-one matrix. We found that DNN with ReLU activation functions and negative weight correlation $c < 0$ are more sensitive to parameter perturbation. Here we examine the sensitivity of relu-DNN to input perturbation by employing the same results developed in~\cite{BoLi2018}. In Fig.~\ref{fig:qL_mf_vs_q0} (c) and (d), we depict the mean field output overlap $q^{L}_{\mathrm{mf}}$ as a function of input overlap $q^{0}$. It is observed that negative weight correlation corresponds to a higher sensitivity to input perturbation, indicating that the relu-DNN with negatively correlated weights generate more complex functions than those with random or positively correlated weights. We conjecture that negative weight correlation develops in very deep ReLU networks when they are trained to performed complex task where a high expressive power is needed, a phenomenon that has been observed in~\cite{Shang2016}.

\begin{figure}
    \centering
    \subfloat[]{\includegraphics[scale=0.38]{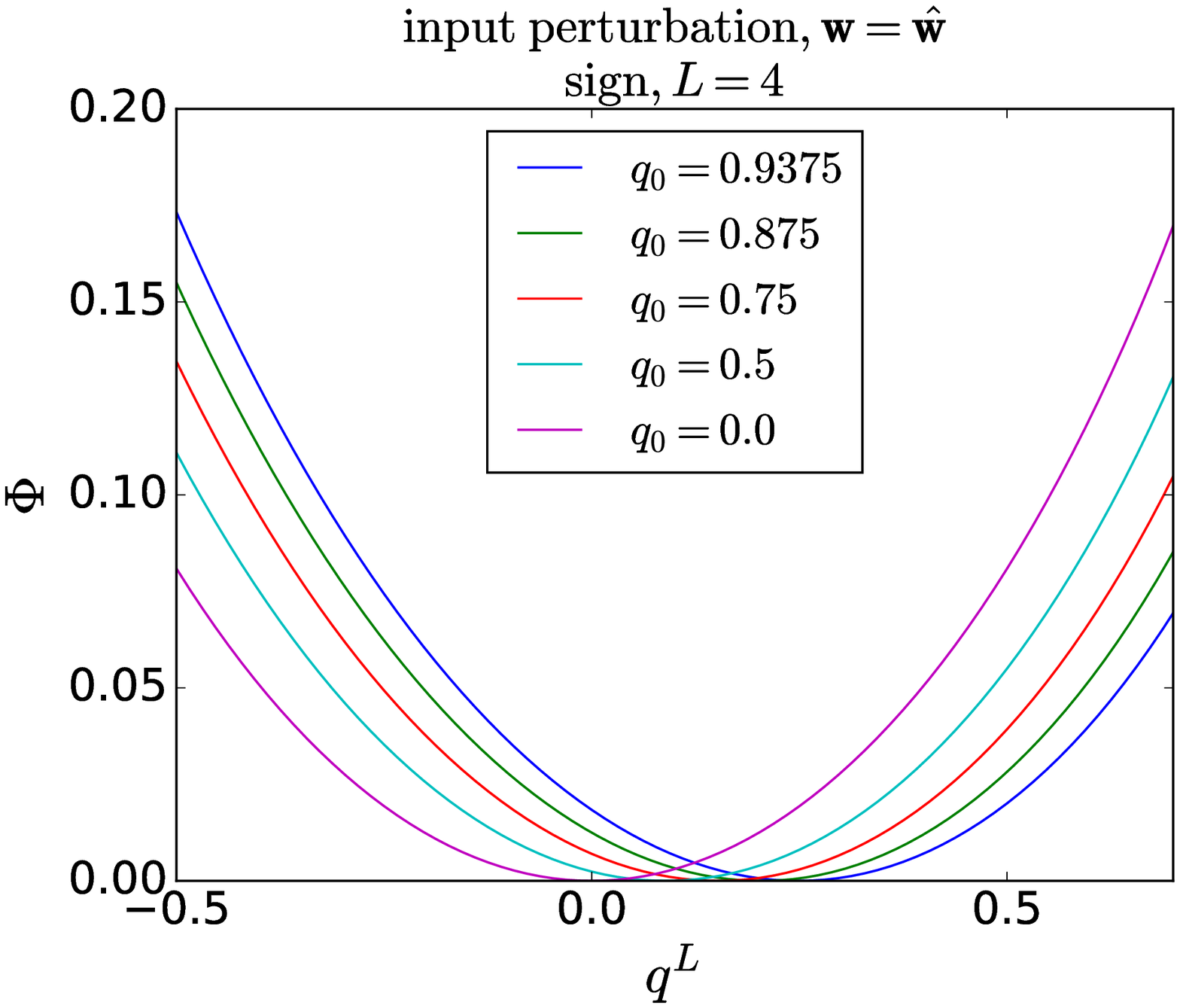}}
    \subfloat[]{\includegraphics[scale=0.38]{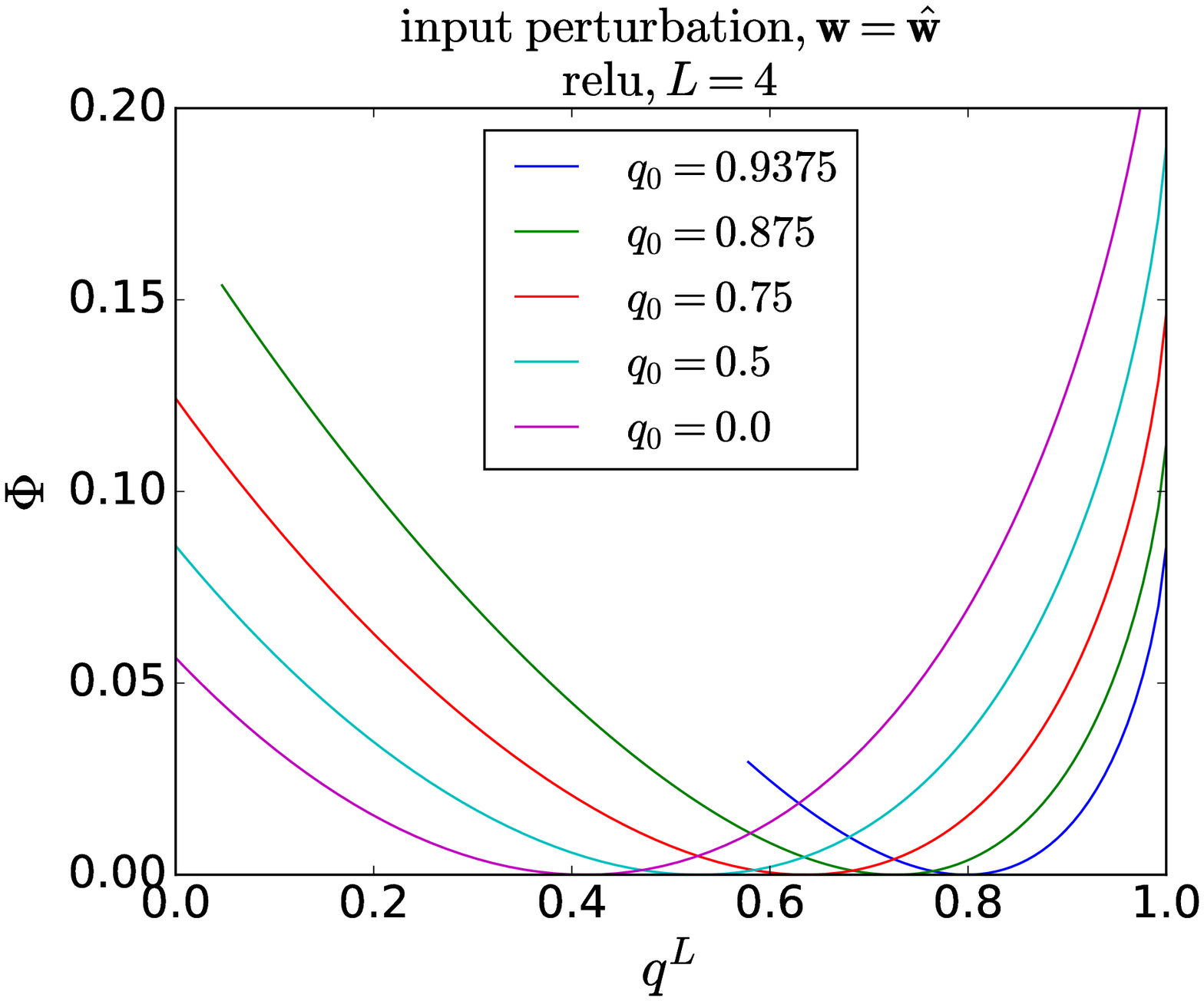}} \\
    \subfloat[]{\includegraphics[scale=0.38]{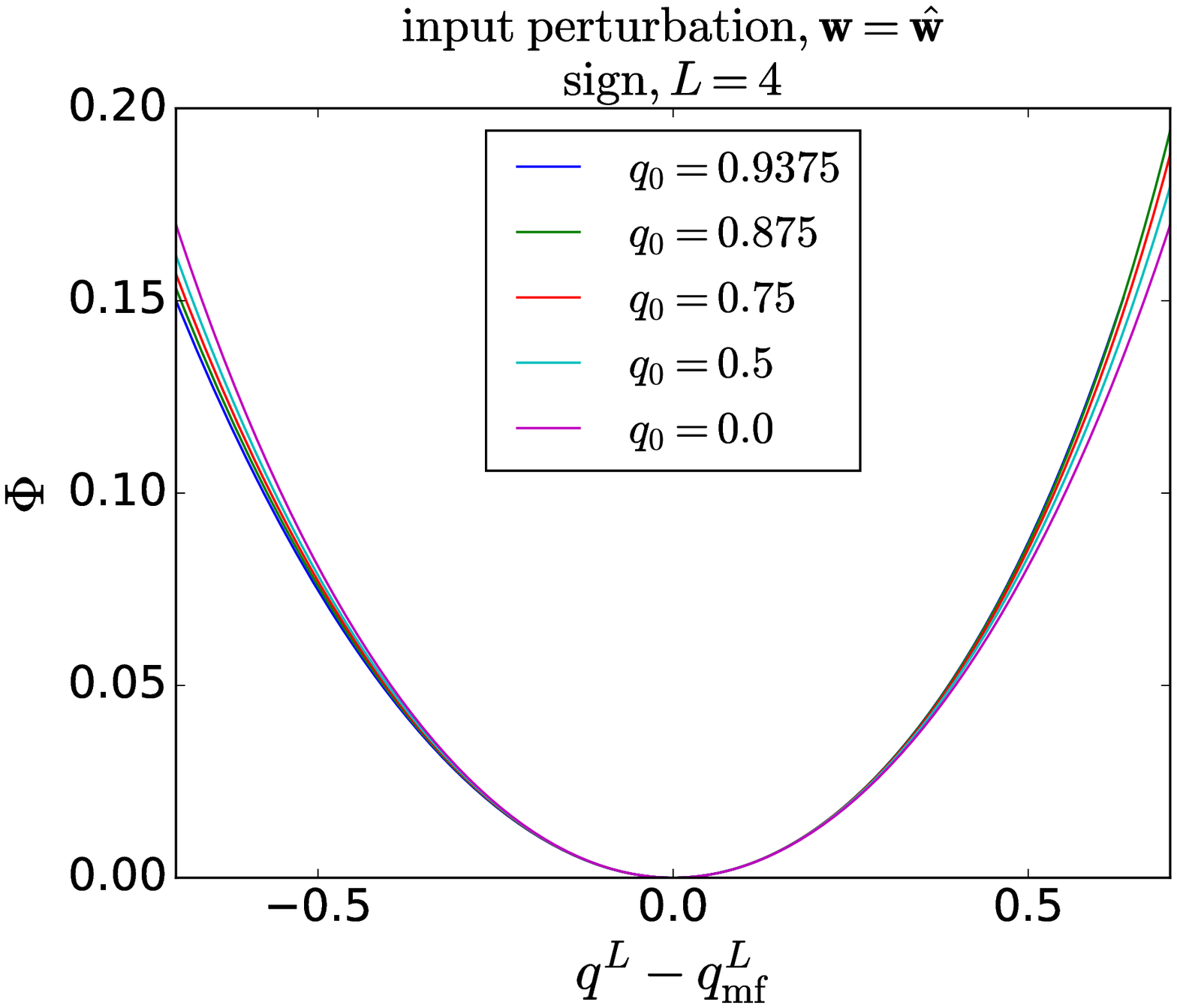}}
    \subfloat[]{\includegraphics[scale=0.38]{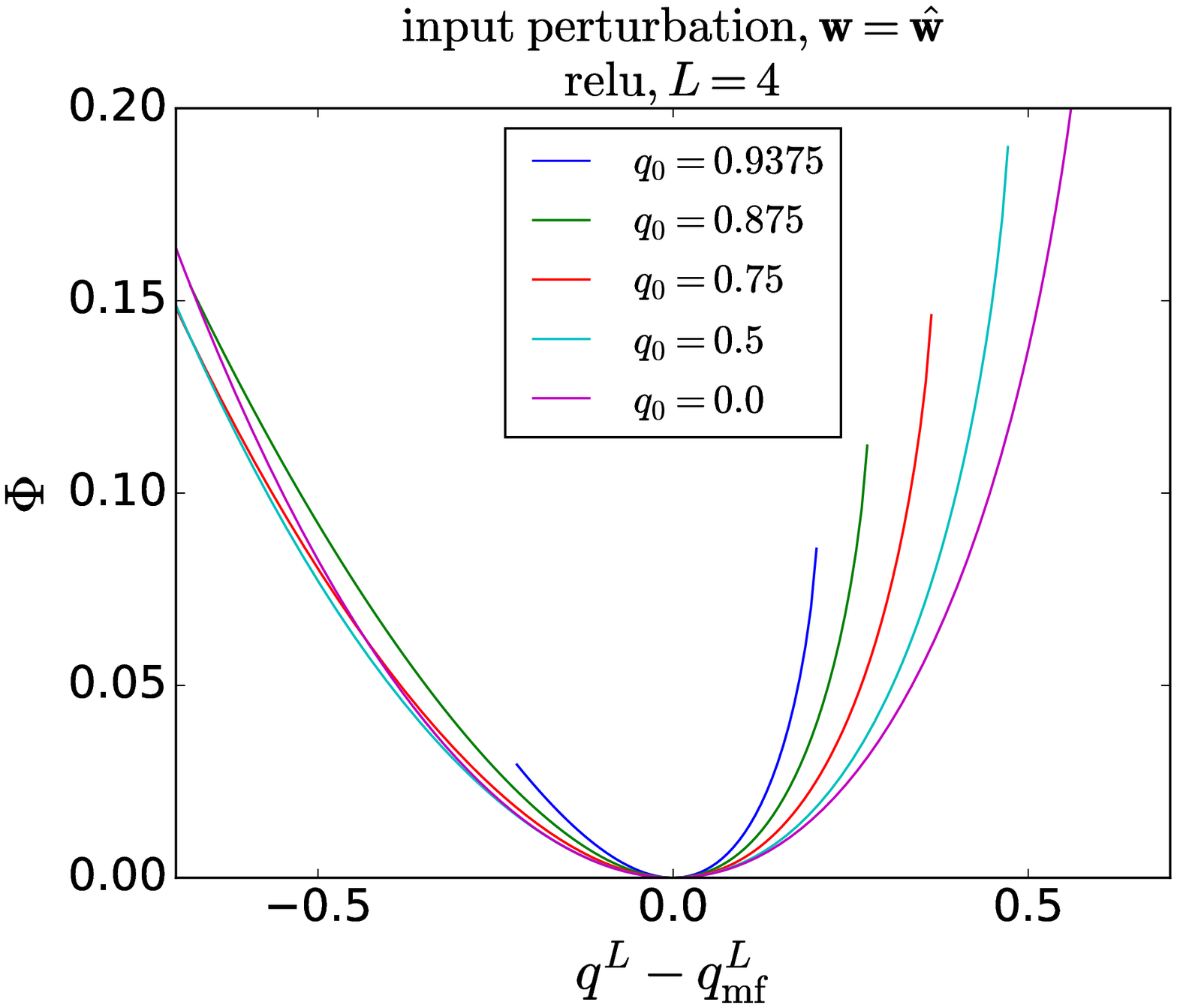}} \\
    \subfloat[]{\includegraphics[scale=0.38]{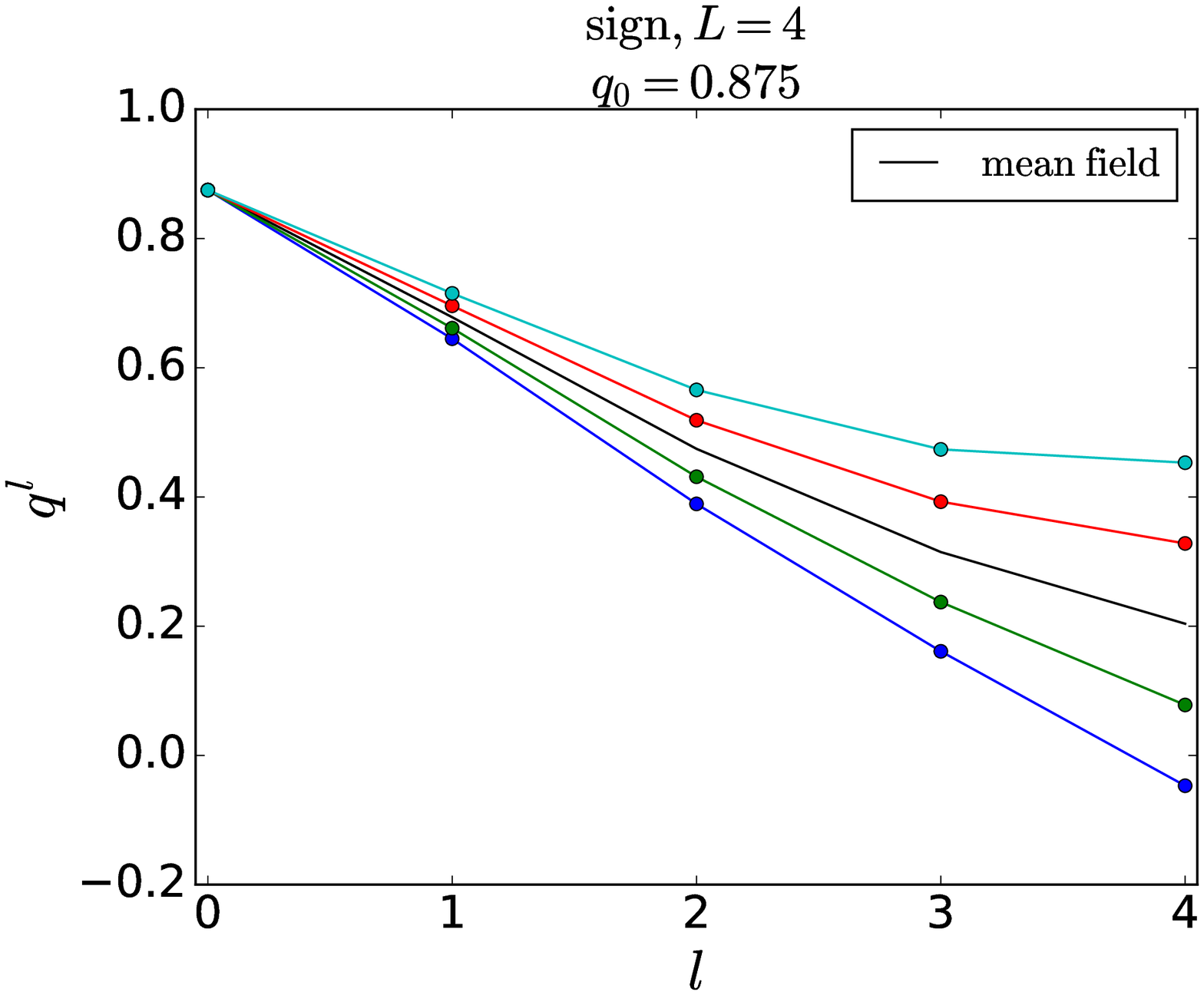}}
    \subfloat[]{\includegraphics[scale=0.38]{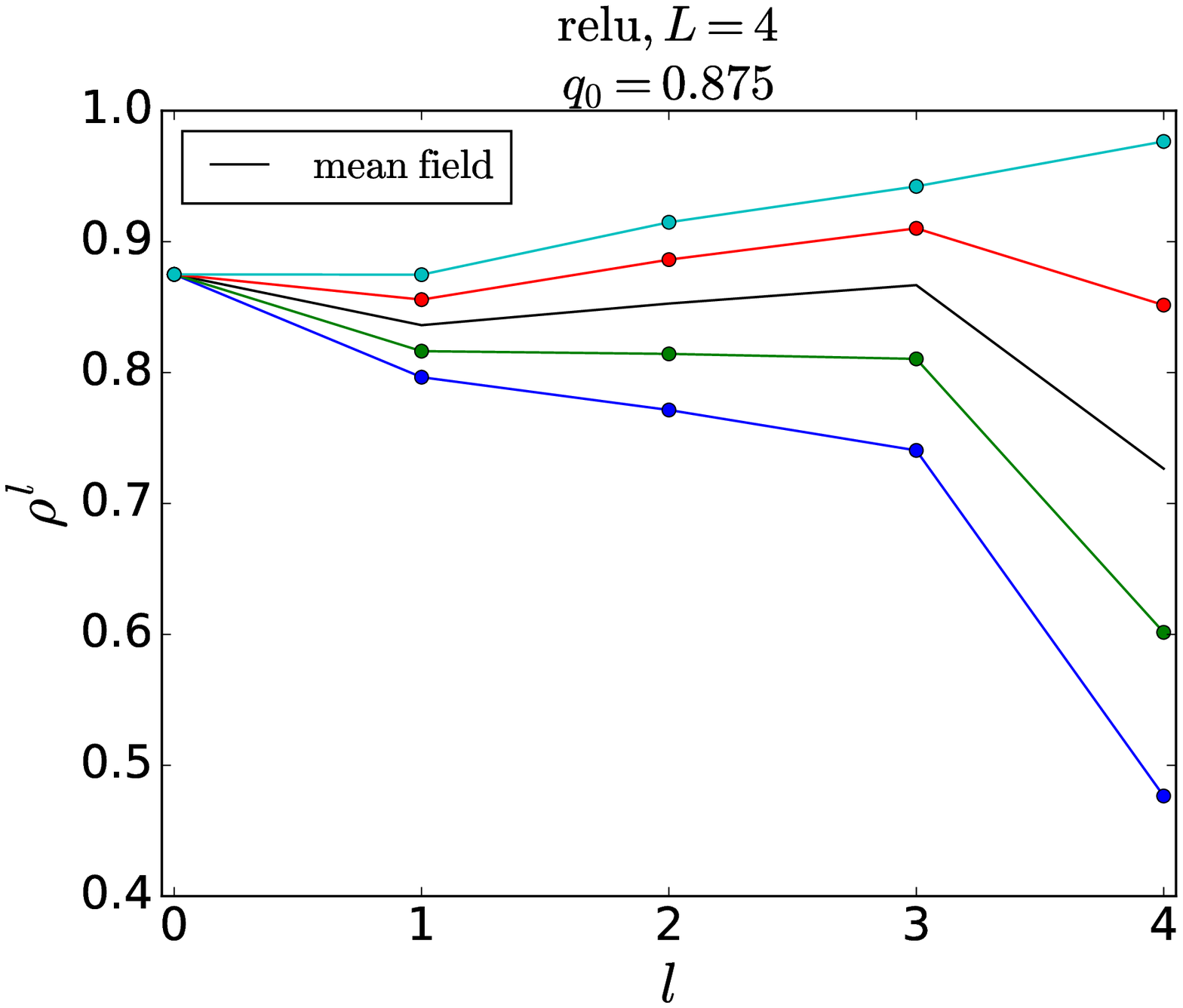}}  
    \caption{Large deviation of output similarity $q^{L}$ under input perturbation where $\boldsymbol{w} = \hat{\boldsymbol{w}}$. Sub-figures (c) and (d) are the same as (a) and (b), except for the shifted $x$-coordinates. (a)(b) $\Phi$ vs $q^{L}$ for sign- and relu-DNN, respectively. (c)(d) $\Phi$ vs $q^{L}-q^{L}_{\mathrm{mf}}$ for sign- and relu-DNN, respectively. (e) The dominant trajectories of overlap $\{ q^{l} \}$ leading to particular deviation in sign-DNN. (f) The dominant trajectories of correlation coefficient $\{ \rho^{l} \}$ leading to particular deviation in relu-DNN. }
    \label{fig:deviation_input}
\end{figure}

In Fig.~\ref{fig:deviation_input}, we further investigate deviations from the typical behaviors in the presence of input perturbations for the specific example with $L=4, \alpha^{l}=1$. The rate functions $\Phi(q^{L})$ depicted in Fig.~\ref{fig:deviation_input}(a)(b) dictate the rate of convergence to the typical behaviors with increasing $N$ by the large deviation principle, for both sign and ReLU activations, respectively. In Fig.~\ref{fig:deviation_input}(c), we observe that the rate functions have similar trends in the vicinity of the mean field solution $q^{L}_{\mathrm{mf}}$ for different levels of input perturbation (corresponding to different $q^{0}$) in sign-DNN, while they are more distinctive in relu-DNN as seen in Fig.~\ref{fig:deviation_input}(d). In relu-DNN, smaller input perturbation (larger $q^{0}$) leads to smaller variance of $q^{L}$ around $q^{L}_{\mathrm{mf}}$. The rate function of relu-DNN is also more asymmetric around $q^{L}_{\mathrm{mf}}$, suggesting that large deviations will be more often observed below $q^{L}_{\mathrm{mf}}$ than above it. This indicates that random relu-DNN of finite size may produce functions that are slightly more complex than what would be expected by the mean field solutions, which remains to be verified.

We also examine the dominant trajectories across layers leading to particular deviations by monitoring the correlations of activations between the two systems across layers. The relevant quantity is the correlation coefficient
\begin{equation}
    \rho^{l} = \frac{ q^{l} - \hat{m}^{l} m^{l} }{ \sqrt{\hat{v}^{l} - (\hat{m}^{l})^2} \sqrt{v^{l} - (m^{l})^2} },
\end{equation}
where the mean activations $\hat{m}^{l}$ and $m^{l}$ are computed by Eq.~(\ref{eq:mean_activation}). We find that sign-DNN satisfy $\hat{m}^{l}=m^{l}=0, \hat{v}^{l}=v^{l}=1$, such that $\rho^{l} = q^{l}$ in this case. The results are shown in Fig.~\ref{fig:deviation_input}(e) and (f), which suggest that the deviations of $q^{L}$ from the typical value $q^{L}_{\mathrm{mf}}$ are mainly contributed by the deviations at later layers.

\begin{figure}
    \centering
    \subfloat[]{\includegraphics[scale=0.38]{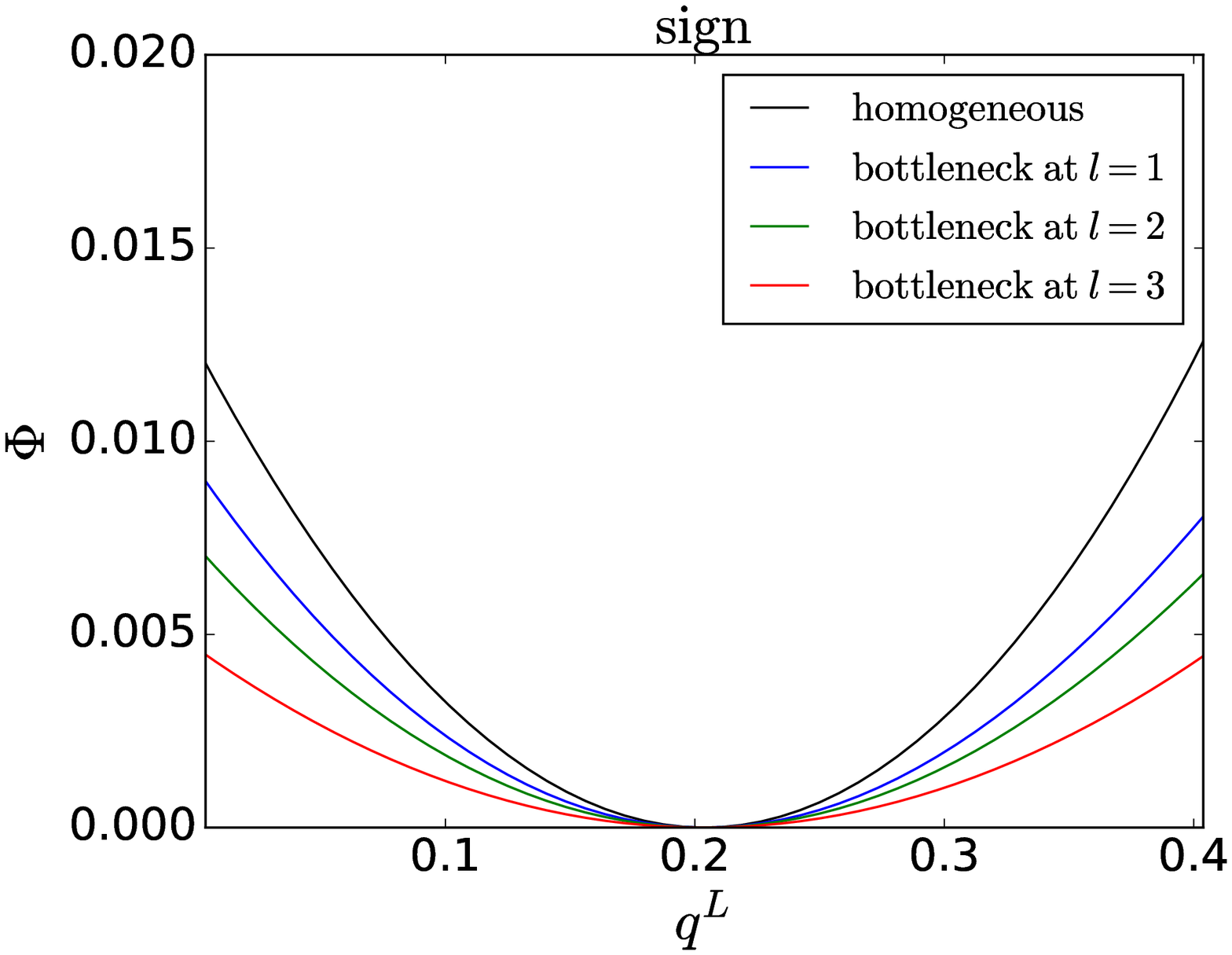}}
    \subfloat[]{\includegraphics[scale=0.38]{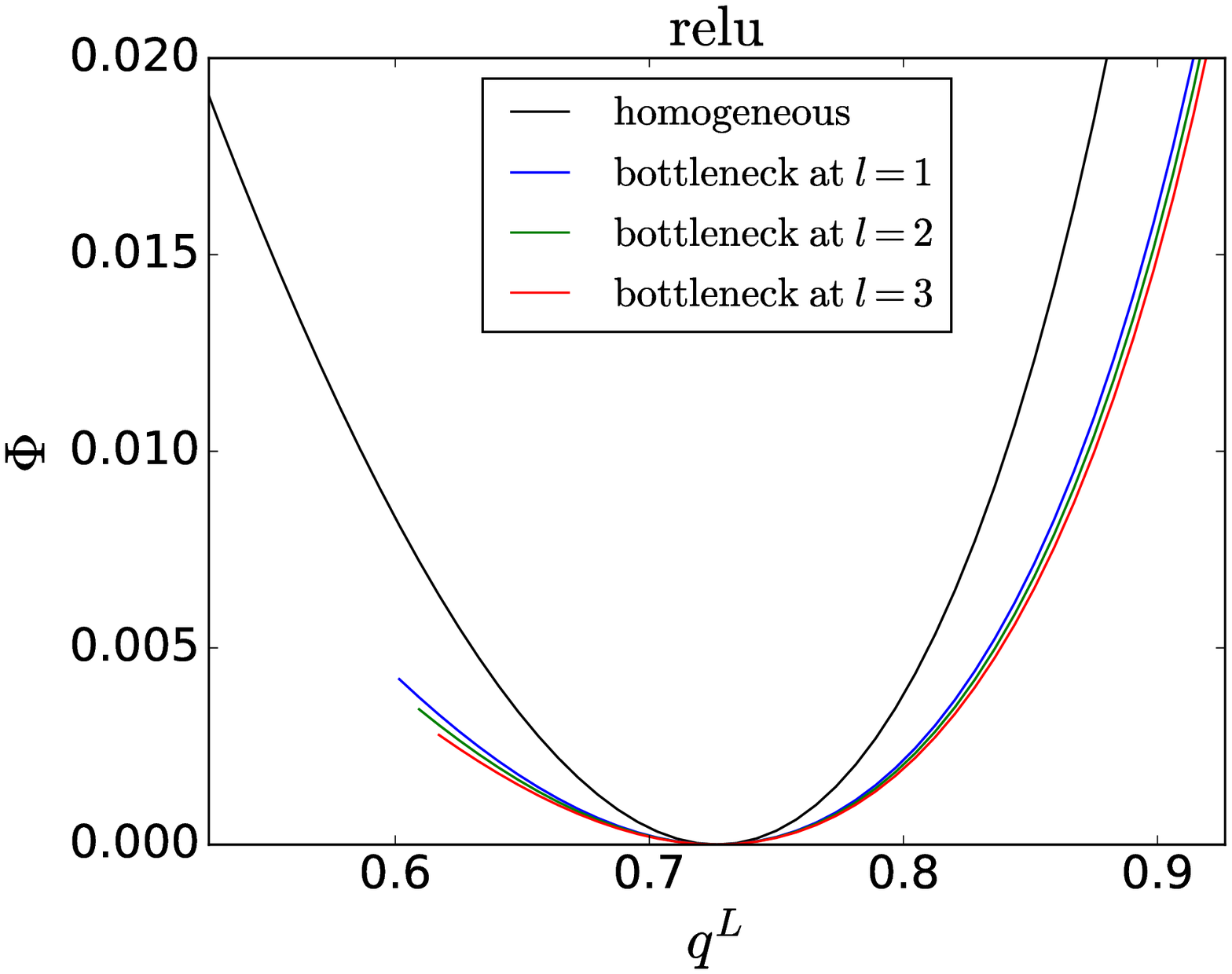}}
    \caption{Effect of a single bottleneck layer on the rate function in the scenario of input perturbation. The bottleneck layer $l'$ has width parameter $\alpha^{l'}=\frac{1}{8}$ while all other layers have $\alpha^{l}=1$. (a) sign-DNN. (b) relu-DNN. }
    \label{fig:bottle_neck}
\end{figure}

Lastly, we investigate the effect of DNN architecture on the deviation. In particular, we consider a single bottleneck layer at a particular hidden layer $l'$ ($0<l'<L$) with $\alpha^{l'}=\frac{1}{8}$ while all other layers satisfy $\alpha^{l}=1,\forall l \neq l'$. Placing the bottleneck at later layer introduces a higher variability of output overlap $q^{L}$ by observing smaller values of the rate function in Fig.~\ref{fig:bottle_neck}; this effect is more prominent in sign-DNN, while it is much less noticeable in relu-DNN.

\section{Discussion}
By utilizing the large deviation theory coupled with the path integral analysis, we derive the sensitivity of finite size random DNN under parameter and input perturbations. Random DNN with sign or ReLU activation function are shown to satisfy the large deviation principle, where the rate functions govern an exponential decay of the deviation to the mean field behaviors as the size of the system increases. We also investigate the effects of weight sparsification and binarization of random DNN, and uncover their equivalence to rotation of weight vector in high dimension. Random DNN with ReLU activation function are found to be robust to these parameter perturbations, which is caused by the low complexity of the corresponding function mappings. Random initializing the weights of ReLU DNN places a prior for simple functions, while they have the capacity to compute more complex functions with specifically trained weights. The next important question is how the networks adapt to perform complex tasks by the training processes.

\ack
BL and DS acknowledge support from the Leverhulme Trust (RPG-2018-092), European Union’s Horizon 2020 research and innovation programme under the Marie Sk{\l}odowska-Curie grant agreement No. 835913. DS acknowledges support from the EPSRC programme grant TRANSNET (EP/R035342/1).

\appendix
\section{Disorder average for weight sparsification} \label{sec:appendix_average_x_sparse}
For network sparsification~(\ref{eq:def_sparse_perturbation}), the disorder average in Eq.~(\ref{eq:PqL_before_average}) can be computed as
\begin{eqnarray}
    \fl \mathbb{E}_{\hat{\boldsymbol{w}}} \prod_{l,i,j} \exp\bigg( \frac{-\im}{\sqrt{N^{l-1}}} \hat{w}^{l}_{ij} \hat{x}^{l}_{i} \hat{s}^{l-1}_{j} \bigg) \bigg[ (1-p^{l}) \exp\bigg( \frac{-\im}{\sqrt{N^{l-1}} \sqrt{1-p^{l}}} \hat{w}^{l}_{ij} x^{l}_{i} s^{l-1}_{j} \bigg) + p^{l} \bigg] \nonumber \\
    \fl = \prod_{l,i,j} \left[ (1-p^{l} )\exp\bigg[ -\frac{\sigma_{w}^2}{2N^{l-1}} \bigg( \hat{x}^{l}_{i} \hat{s}^{l-1}_{j} + x^{l}_{i} s^{l-1}_{j}/\sqrt{1-p^{l}} \bigg)^2 \bigg] + p^{l} \exp\bigg[ -\frac{\sigma_{w}^2}{2N^{l-1}} \bigg( \hat{x}^{l}_{i} \hat{s}^{l-1}_{j} \bigg)^2 \bigg] \right] \nonumber \\
    \fl = \prod_{l,i,j} \left\{ (1-p^{l} )\bigg[ 1 -\frac{\sigma_{w}^2}{2N^{l-1}} \bigg( \hat{x}^{l}_{i} \hat{s}^{l-1}_{j} + x^{l}_{i} s^{l-1}_{j}/\sqrt{1-p^{l}} \bigg)^2 \bigg] \right. \nonumber \\
    \fl \qquad \quad \left. + p^{l} \bigg[ 1 -\frac{\sigma_{w}^2}{2N^{l-1}} \bigg( \hat{x}^{l}_{i} \hat{s}^{l-1}_{j} \bigg)^2 \bigg] + O\bigg( \frac{1}{(N^{l-1})^2} \bigg) \right\} \nonumber \\
    \fl \approx \prod_{l,i,j} \left\{ 1- \frac{\sigma_{w}^{2}}{N^{l-1}} \bigg[ \frac{1}{2}(\hat{x}^{l}_{i})^2 (\hat{s}^{l-1}_{j})^2 + \frac{1}{2}(x^{l}_{i})^2 (s^{l-1}_{j})^2 + \sqrt{1-p^{l}} (\hat{x}^{l}_{i} x^{l}_{i}) (\hat{s}^{l-1}_{j} s^{l-1}_{j}) \bigg] \right\} \nonumber \\
    \fl \approx \prod_{l,i} \exp \left\{ -\sigma_{w}^{2} \bigg[\frac{1}{2}(\hat{x}_{i}^{l})^{2} \frac{\sum_{j} (\hat{s}^{l-1}_{j})^2}{N^{l-1}} +\frac{1}{2}(x_{i}^{l})^{2} \frac{\sum_{j} (s^{l-1}_{j})^2}{N^{l-1}} +\sqrt{1-p^{l}}\hat{x}_{i}^{l}x_{i}^{l} \frac{\sum_{j} \hat{s}^{l-1}_{j} s^{l-1}_{j}}{N^{l-1}} \bigg] \right\},
\end{eqnarray}
where we have made use of the large $N^{l}$ approximation.

\section{Disorder average for weight binarization} \label{sec:appendix_average_x_binary}
For weight binarization in~(\ref{eq:def_binary_perturbation}), the disorder average in Eq.~(\ref{eq:PqL_before_average}) can be computed as
\begin{eqnarray}
    \fl \mathbb{E}_{\hat{\boldsymbol{w}}} \prod_{l,i,j} \exp\bigg[ \frac{-\im}{\sqrt{N^{l-1}}} \bigg( \hat{w}^{l}_{ij} \hat{x}^{l}_{i} \hat{s}^{l-1}_{j} + \mathrm{sgn}(\hat{w}^{l}_{ij}) \sigma_{w} x^{l}_{i} s^{l-1}_{j} \bigg) \bigg] \nonumber \\
    \fl = \prod_{l,i,j} \left\{ \int_{-\infty}^{0} \diff \hat{w}^{l}_{ij} \mathcal{N}(\hat{w}^{l}_{ij}|0, \sigma_{w}^2) \exp\bigg[ \frac{-\im}{\sqrt{N^{l-1}}} \bigg( \hat{w}^{l}_{ij} \hat{x}^{l}_{i} \hat{s}^{l-1}_{j} - \sigma_{w} x^{l}_{i} s^{l-1}_{j} \bigg) \bigg] \right. \nonumber \\
    \fl \qquad \quad \left. + \int_{0}^{\infty} \diff \hat{w}^{l}_{ij} \mathcal{N}(\hat{w}^{l}_{ij}|0, \sigma_{w}^2) \exp\bigg[ \frac{-\im}{\sqrt{N^{l-1}}} \bigg( \hat{w}^{l}_{ij} \hat{x}^{l}_{i} \hat{s}^{l-1}_{j} + \sigma_{w} x^{l}_{i} s^{l-1}_{j} \bigg) \bigg] \right\} \nonumber \\
    \fl = \prod_{l,i,j} \exp\bigg[ - \frac{\sigma_{w}^{2}}{2N^{l-1}} (\hat{x}^{l}_{i})^2 (\hat{s}^{l-1}_{j})^2 \bigg] \frac{1}{2} \left\{ \bigg[ 1+ \mathrm{erf}\bigg( \frac{\im \hat{x}^{l}_{j} \hat{s}^{l-1}_{j} \sigma_{w} }{2\sqrt{N^{l-1}}} \bigg) \bigg] \exp\bigg( \frac{\im x^{l}_{i} s^{l-1}_{j} \sigma_{w} }{\sqrt{N^{l-1}}} \bigg) \right. \nonumber \\
    \fl \qquad \quad \left. + \bigg[ 1 - \mathrm{erf}\bigg( \frac{\im \hat{x}^{l}_{j} \hat{s}^{l-1}_{j} \sigma_{w} }{\sqrt{2N^{l-1}}} \bigg) \bigg] \exp\bigg( \frac{-\im x^{l}_{i} s^{l-1}_{j} \sigma_{w} }{\sqrt{N^{l-1}}} \bigg) \right\} \nonumber \\
    \fl  = \prod_{l,i,j} \exp\bigg[ - \frac{\sigma_{w}^{2}}{2N^{l-1}} (\hat{x}^{l}_{i})^2 (\hat{s}^{l-1}_{j})^2 \bigg] \frac{1}{2} \left\{ \bigg( 1+ \frac{2}{\sqrt{\pi}} \frac{\im \hat{x}^{l}_{j} \hat{s}^{l-1}_{j} \sigma_{w} }{\sqrt{2N^{l-1}}} \bigg) \bigg[ 1 + \frac{\im x^{l}_{i} s^{l-1}_{j} \sigma_{w} }{\sqrt{N^{l-1}}} - \frac{1}{2}  \frac{ (x^{l}_{i} s^{l-1}_{j} \sigma_{w})^2 }{N^{l-1}} \bigg] \right. \nonumber \\
    \fl \qquad \quad \left. + \bigg( 1 - \frac{2}{\sqrt{\pi}} \frac{\im \hat{x}^{l}_{j} \hat{s}^{l-1}_{j} \sigma_{w} }{\sqrt{2N^{l-1}}} \bigg) \bigg[ 1 - \frac{\im x^{l}_{i} s^{l-1}_{j} \sigma_{w} }{\sqrt{N^{l-1}}} - \frac{1}{2}  \frac{ (x^{l}_{i} s^{l-1}_{j} \sigma_{w} )^2 }{N^{l-1}} \bigg] + O\bigg( \frac{1}{(N^{l-1})^2} \bigg) \right\} \nonumber \\
    \fl \approx \prod_{l,i,j} \exp\bigg[ - \frac{\sigma_{w}^{2}}{2N^{l-1}} (\hat{x}^{l}_{i})^2 (\hat{s}^{l-1}_{j})^2 \bigg] \left\{ 1- \frac{\sigma_{w}^{2}}{N^{l-1}} \bigg[ \frac{1}{2}(x^{l}_{i})^2 (s^{l-1}_{j})^2 + \sqrt{\frac{2}{\pi}} (\hat{x}^{l}_{i} x^{l}_{i}) (\hat{s}^{l-1}_{j} s^{l-1}_{j}) \bigg] \right\} \nonumber \\
    \fl \approx \prod_{l,i} \exp \left\{ -\sigma_{w}^{2} \bigg[\frac{1}{2}(\hat{x}_{i}^{l})^{2} \frac{\sum_{j} (\hat{s}^{l-1}_{j})^2}{N^{l-1}} +\frac{1}{2}(x_{i}^{l})^{2} \frac{\sum_{j} (s^{l-1}_{j})^2}{N^{l-1}} +\sqrt{\frac{2}{\pi}}\hat{x}_{i}^{l}x_{i}^{l} \frac{\sum_{j} \hat{s}^{l-1}_{j} s^{l-1}_{j}}{N^{l-1}} \bigg] \right\},
\end{eqnarray}
where the large $N^{l}$ approximation has been employed.

\section{Large deviation in the multiple-pattern scenario}\label{sec:appendix_multiple_input}
Consider function similarity estimated for multiple patterns
\begin{eqnarray}
    \tilde{q}^{L} & = \frac{1}{M} \sum_{\mu=1}^{M} \bigg( \frac{1}{N^{L}} \sum_{i=1}^{N^{L}} \hat{s}^{L,\mu}_{i} s^{L,\mu}_{i} \bigg) =: \frac{1}{M} \sum_{\mu=1}^{M} q^{L,\mu}
\end{eqnarray}
where $\hat{s}^{L,\mu}_{i}(\boldsymbol{\hat{s}}^{0,\mu})$ is the $i$th output of the reference network with the $\mu$th input $\boldsymbol{\hat{s}}^{0,\mu}$ drawn independently and identically from the input distribution $P(\boldsymbol{s}^{0})$. In the small fluctuation regime, where each $q^{L,\mu}$ is close to the mean field solution $q^{L}_{\mathrm{mf}}$, we have $I(q^{L,\mu}) \approx 1/2I''(q^{L}_{\mathrm{mf}}) (q^{L,\mu} - q^{L}_{\mathrm{mf}})^2$ (both $I(q^{L}_{\mathrm{mf}})$ and $I'(q^{L}_{\mathrm{mf}})$ vanish~\cite{Touchette2009}), i.e., $P(q^{L,\mu})$ can be approximated by a Gaussian density \begin{equation}
    P(q^{L,\mu}) \sim \exp \bigg( -\frac{N}{2}I''(q^{L}_{\mathrm{mf}}) (q^{L,\mu} - q^{L}_{\mathrm{mf}})^2 \bigg),
\end{equation}
where the corresponding variance is $1/(NI''(q^{L}_{\mathrm{mf}}))$. Since the $M$ inputs are independent, we also assume the outputs are also approximately independent (which holds in sign-DNN but does not necessary for relu-DNN since ReLU non-linearity can induce correlations among variables), such that the variance of $\tilde{q}^{L}$ is $1/(MNI''(q^{L}_{\mathrm{mf}}))$. Therefore, in the vicinity of $q^{L}_{\mathrm{mf}}$ we have
\begin{equation}
    P(\tilde{q}^{L}) \sim \exp \bigg( -\frac{MN}{2}I''(q^{L}_{\mathrm{mf}}) (\tilde{q}^{L} - q^{L}_{\mathrm{mf}})^2 \bigg),
\end{equation}
implying that the corresponding rate function differs from the one with single pattern by a factor of $M$.

More formally, one can directly compute the probability density $P({\tilde{q}^L})$ as
\begin{eqnarray}
    \fl P({\tilde{q}^L}) = \left\langle \delta \bigg( \frac{1}{MN^{L}} \sum_{\mu,i} \hat{s}^{L,\mu}_{i} s^{L,\mu}_{i} - \tilde{q}^L \bigg) \right\rangle \nonumber \\
    \fl = \mathbb{E}_{\hat{\boldsymbol{w}}, \boldsymbol{w}} \mathrm{Tr}_{\hat{\boldsymbol{s}}, \boldsymbol{s}} \delta \bigg( \frac{1}{MN^{L}} \sum_{\mu,i} \hat{s}^{L,\mu}_{i} s^{L,\mu}_{i} - \tilde{q}^L \bigg) \prod_{\mu, i} P(\hat{s}^{0,\mu}_{i}) \delta_{s_{i}^{0,\mu},\hat{s}_{i}^{0,\mu}} \int \prod_{\mu,l,i} \frac{\diff \hat{h}^{l,\mu}_{i} \diff \hat{x}^{l,\mu}_{i} }{2\pi} \frac{\diff h^{l,\mu}_{i} \diff x^{l,\mu}_{i} }{2\pi} \nonumber \\
    \fl \quad \times \exp \left[ \sum_{\mu,l,i} \bigg( \log P(\hat{s}^{l,\mu}_{i}|\hat{h}^{l,\mu}_{i}) + \log P(s^{l,\mu}_{i}|h^{l,\mu}_{i}) + \im \hat{x}^{l,\mu}_{i} \hat{h}^{l,\mu}_{i} + \im x^{l,\mu}_{i} h^{l,\mu}_{i} \bigg) \right]  \nonumber \\
    \fl \quad \times \exp \left[ - \sum_{\mu,l} \frac{\im}{\sqrt{N^{l-1}}} \sum_{i,j} \bigg( \hat{w}^{l}_{ij} \hat{x}^{l,\mu}_{i} \hat{s}^{l-1,\mu}_{j} + w^{l}_{ij} x^{l,\mu}_{i} s^{l-1,\mu}_{j} \bigg) \right]. \label{eq:PqL_multi_pattern}
\end{eqnarray}
Since the weights $\{ \hat{w}^{l}_{ij}, w^{l}_{ij} \}$ are shared among the $M$ patterns, average over these variables on the last line of Eq.~(\ref{eq:PqL_multi_pattern}) leads to coupling between patterns on the pre-activation fields
\begin{eqnarray}
    \fl \prod_{l,i} \exp \left\{ -\sigma_{w}^{2} \sum_{\mu, \nu} \bigg[\frac{1}{2}\hat{x}_{i}^{l,\mu} \hat{x}_{i}^{l,\nu} \frac{1}{N^{l-1}} \sum_{j} \hat{s}^{l-1,\mu}_{j} \hat{s}^{l-1,\nu}_{j} +\frac{1}{2} x_{i}^{l,\mu} x_{i}^{l,\nu} \frac{1}{N^{l-1}} \sum_{j} s^{l-1,\mu}_{j} s^{l-1,\nu}_{j} \right. \nonumber \\
    \fl \qquad \qquad \left. + \sqrt{1-(\eta^{l})^2}\hat{x}_{i}^{l,\mu}x_{i}^{l,\nu} \frac{1}{N^{l-1}} \sum_{j} \hat{s}^{l-1,\mu}_{j} s^{l-1,\nu}_{j} \bigg] \right\}. \label{eq:disorder_average_x_multi_pattern}
\end{eqnarray}
By introducing the following overlap matrices as macroscopic order parameters
\begin{equation}
    \fl q^{l,\mu\nu} = \frac{1}{N^{l}} \sum_{j} \hat{s}^{l,\mu}_{j} s^{l,\nu}_{j}, \quad \hat{v}^{l,\mu\nu} = \frac{1}{N^{l}} \sum_{j} \hat{s}^{l,\mu}_{j} \hat{s}^{l,\nu}_{j}, \quad v^{l,\mu\nu} = \frac{1}{N^{l}} \sum_{j} s^{l,\mu}_{j} s^{l,\nu}_{j},
\end{equation}
Eq.~(\ref{eq:PqL_multi_pattern}) can be factorized over sites as before. However, we have $O(LM^2)$ order parameters here, while there are only $O(L)$ order parameters in the single pattern case. To further simplify the calculation, we assume a symmetric structure of the cross-pattern overlaps at the saddle point $q^{l,\mu\nu} = q^{l, \parallel}\delta_{\mu\nu} + q^{l,\perp}(1-\delta_{\mu\nu})$, where $q^{l, \parallel}, q^{l, \perp}$ are the diagonal and off-diagonal matrix elements respectively. Under this assumption, one can in principle evaluate the integral in~\ref{eq:PqL_multi_pattern}, but the resulting calculation becomes rather involved.

Alternatively, since the $M$ input patterns are independent, we expect the diagonal elements of the matrix $q^{l,\mu\nu}$ to be larger than the off-diagonal elements (sum of correlated variables v.s. sum of random variables). In particular, for sign activation we expect $q^{l,\parallel}\sim O(1), q^{l,\perp}\sim O(\frac{1}{\sqrt{N^{l}}})$ since $q^{l,\perp}$ involves a summation over weakly correlated positive and negative numbers. We therefore approximate the summation $\sum_{\mu\nu}[...]$ in the exponential of Eq.~(\ref{eq:disorder_average_x_multi_pattern}) by $\sum_{\mu = \nu}[...]$, which yields $MN^{l}$ un-coupled identical integrals at each layer $N^{l}$. It eventually leads to the rate function of multiple-pattern overlap $\tilde{q}^{L}$ as $I(\tilde{q}^{L}) \approx M\Phi(\boldsymbol{Q}^*,\boldsymbol{q}^*,...|\tilde{q}^{L})$, where $\Phi(\boldsymbol{Q}^*,\boldsymbol{q}^*,...|q^{L})$ is the rate function of the single-pattern overlap $q^{L}$. While the off-diagonal elements of $q^{l,\mu\nu}$ have smaller values, there are more of these terms ($M(M-1)$ off-diagonal terms compared to $M$ diagonal terms in the summation $\sum_{\mu\nu}[...]$ in the exponential of Eq.~(\ref{eq:disorder_average_x_multi_pattern})), so we expect the above approximation to hold only for small $M$. The above argument may fail for ReLU activation, since $\hat{s}^{l,\mu}_{j}, s^{l,\mu}_{j}$ are always positive, and therefore $q^{l,\perp}\sim O(1)$.

\begin{figure}
    \centering
    \subfloat[]{\includegraphics[scale=0.25]{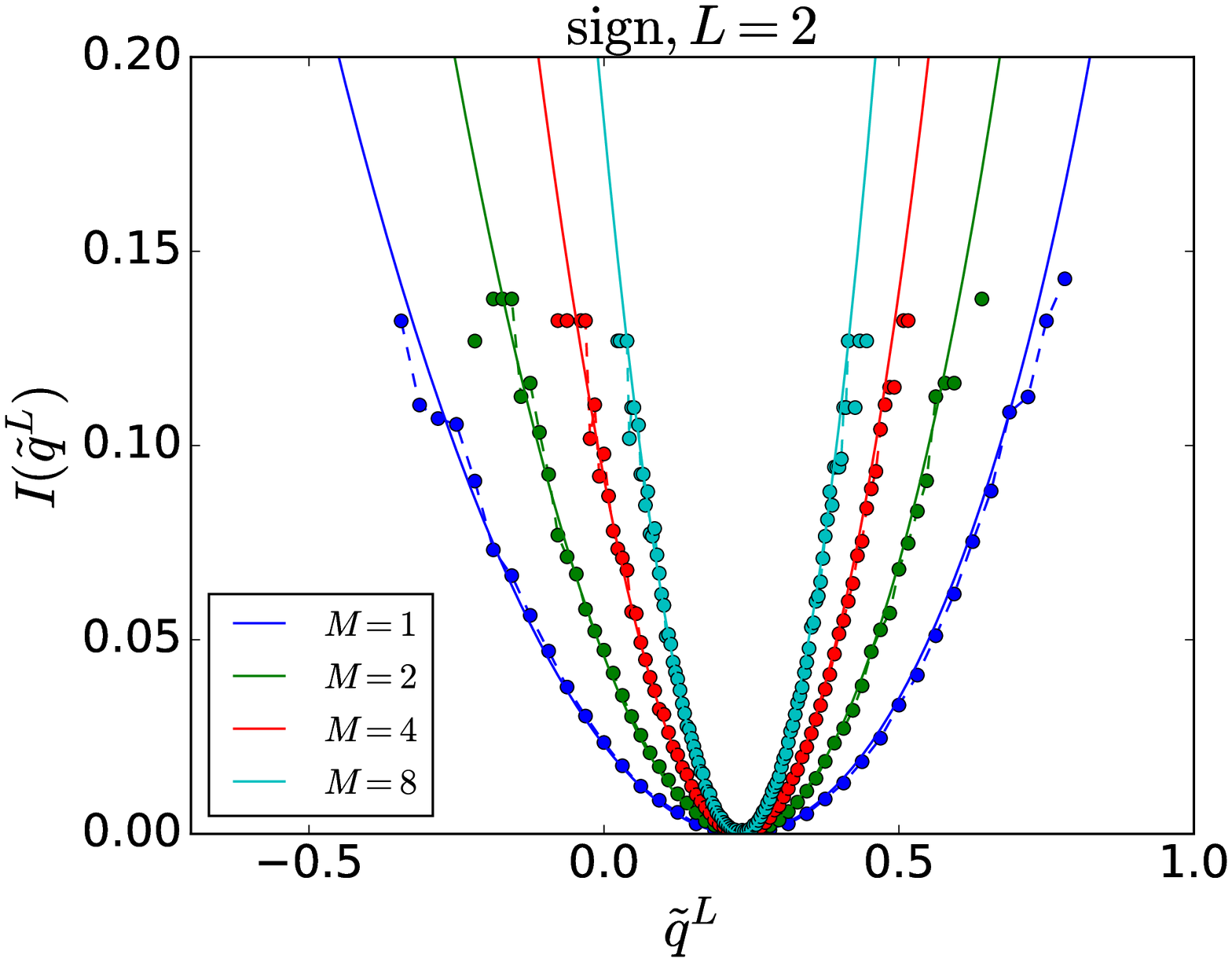}}
    \subfloat[]{\includegraphics[scale=0.25]{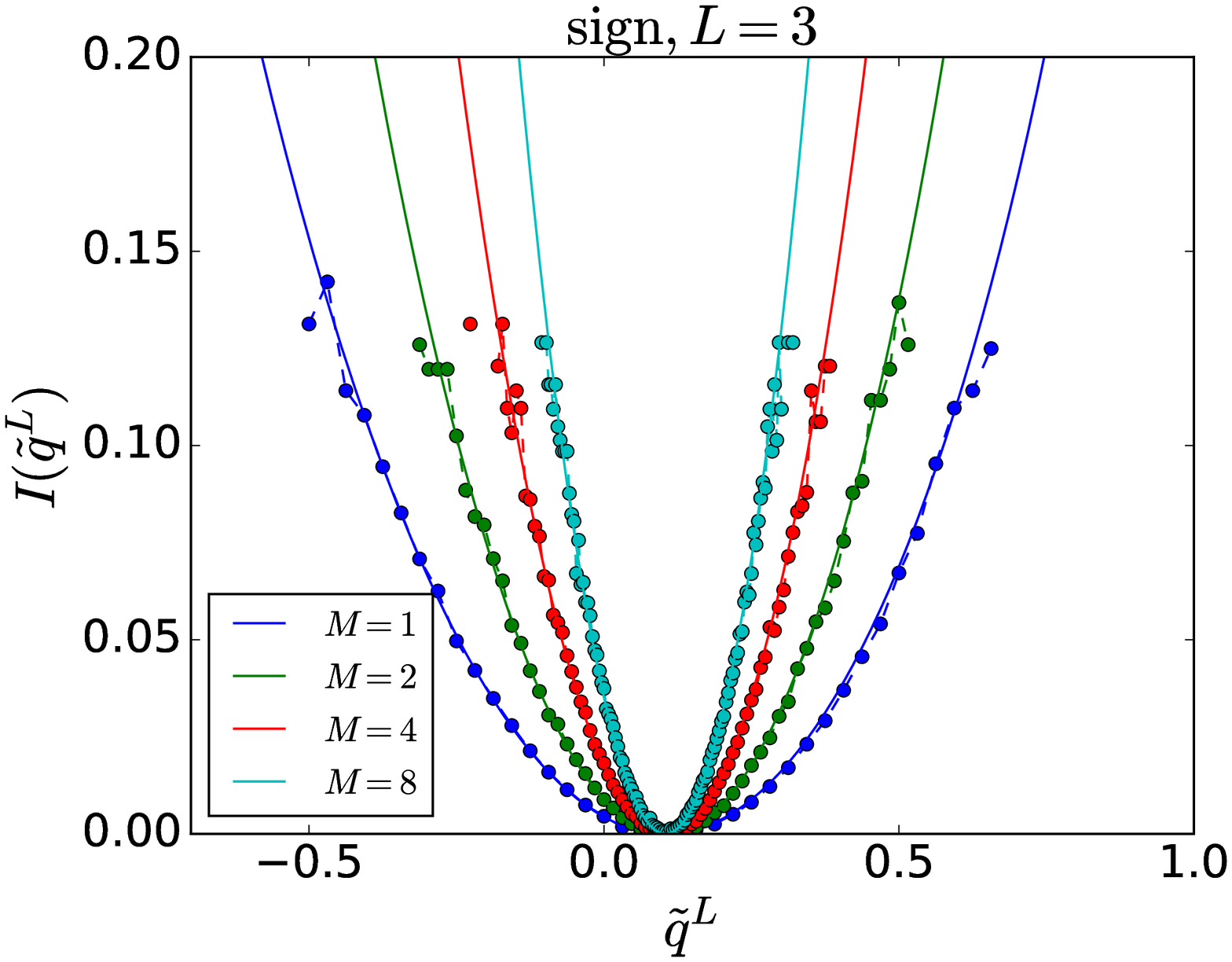}}
    \subfloat[]{\includegraphics[scale=0.25]{Phi_vs_tildeqL_sign_sparse_L4.eps}} \\
    \subfloat[]{\includegraphics[scale=0.25]{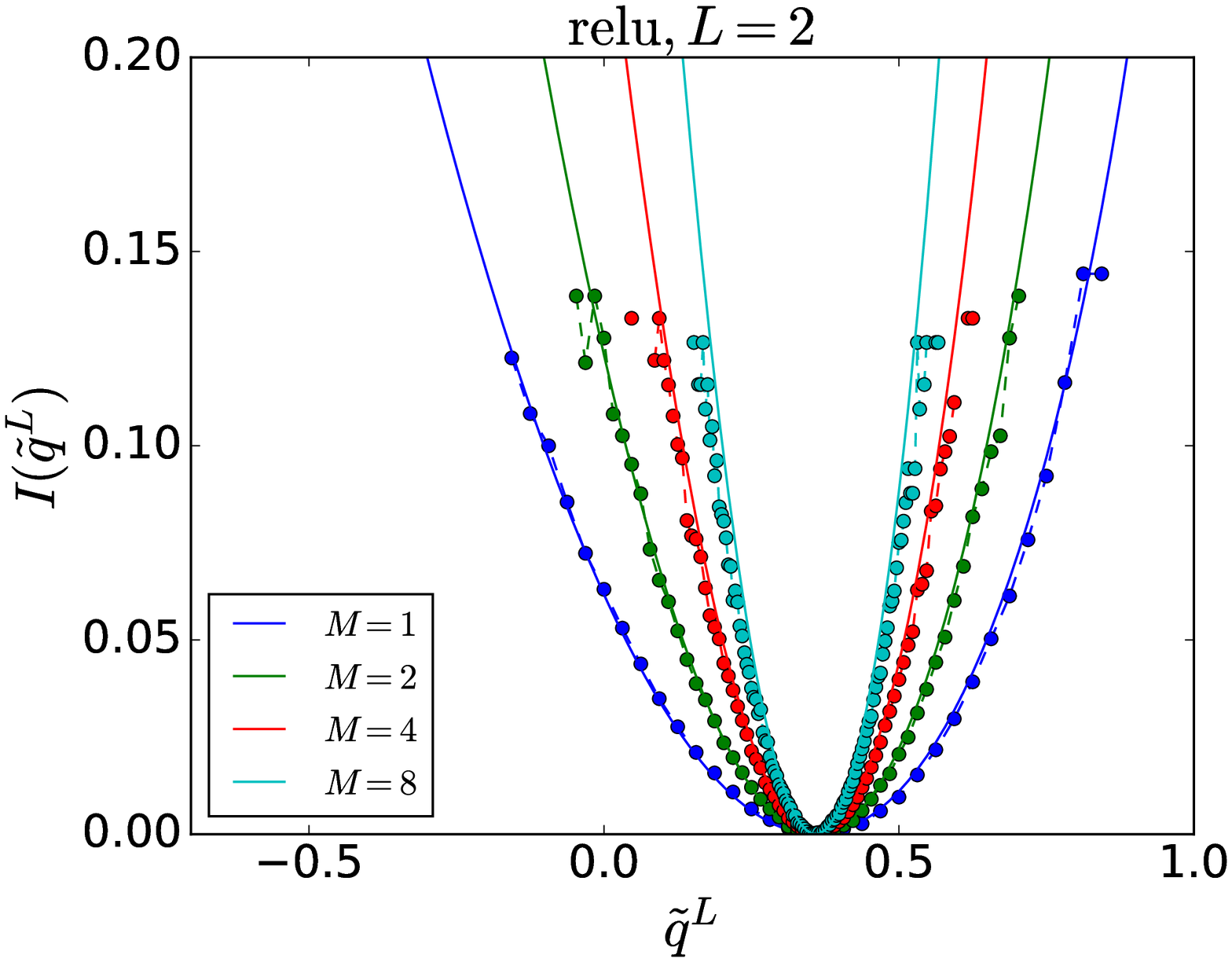}}
    \subfloat[]{\includegraphics[scale=0.25]{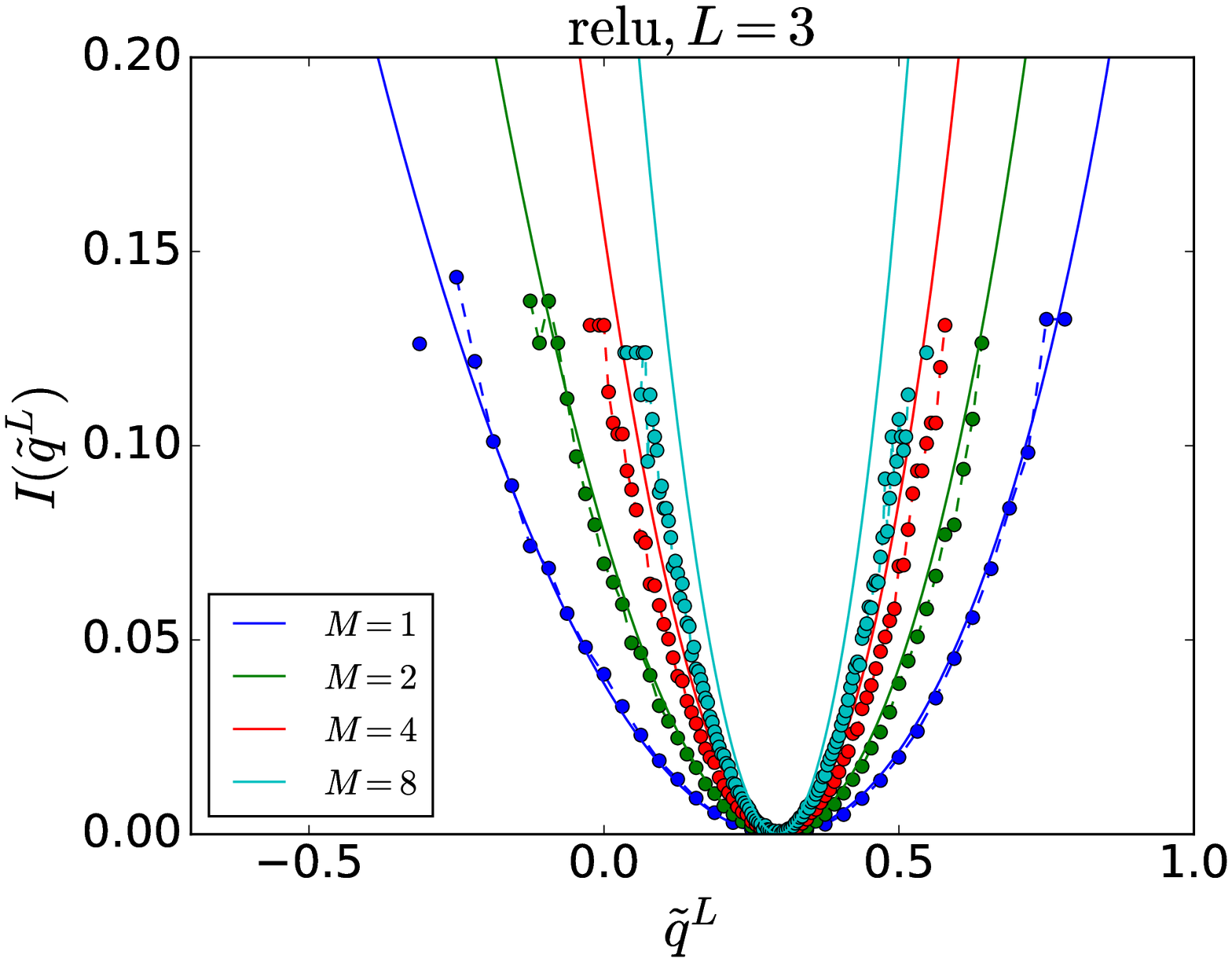}}
    \subfloat[]{\includegraphics[scale=0.25]{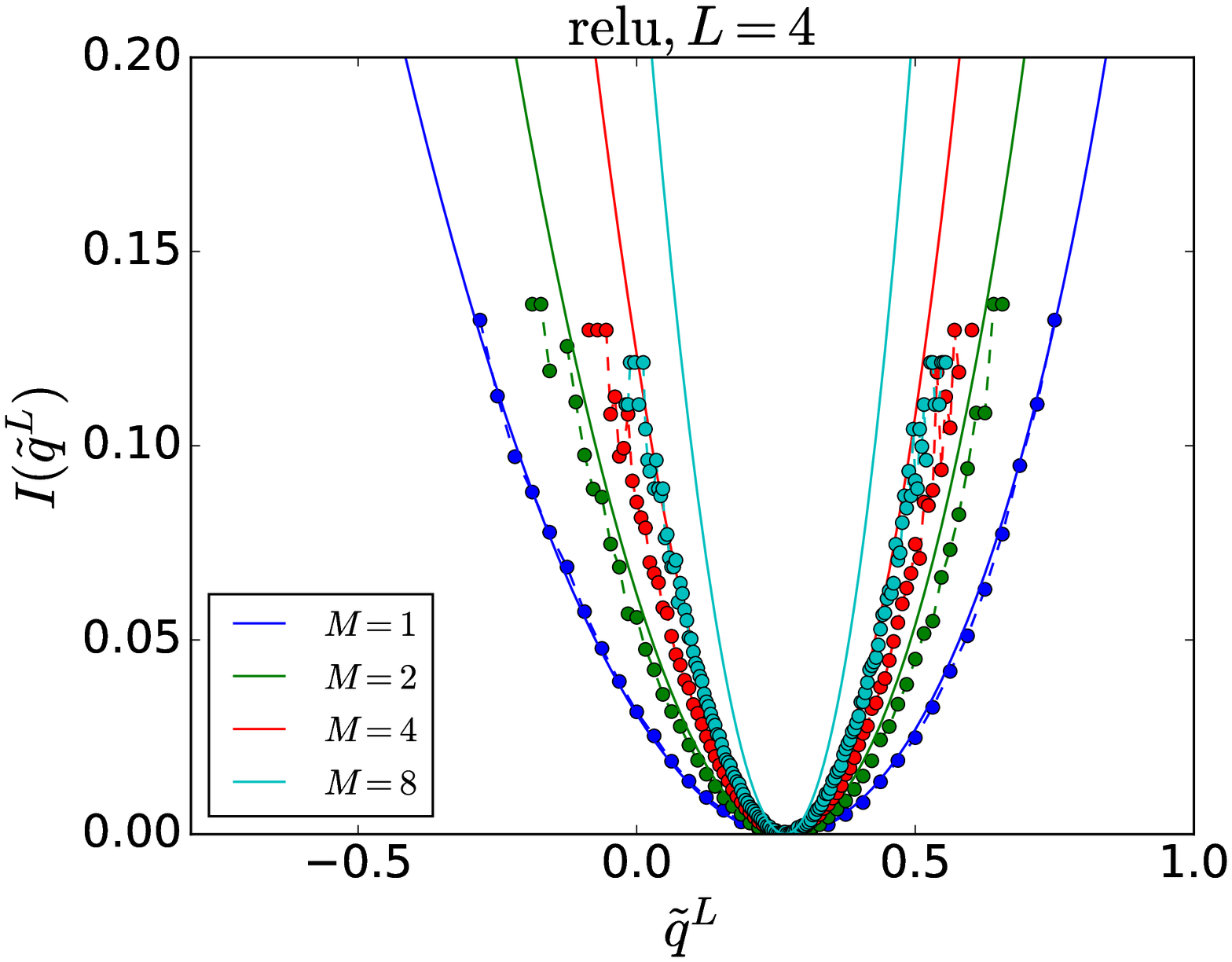}}
    \caption{The rate function $I(\tilde{q}^{L})$ of output overlap $\tilde{q}^{L}$ defined for $M$ patterns and DNN with different activation functions and system depths, in the scenario of weight sparsification with disconnection probability $p^{l}=1/2$. Solid lines correspond to theoretical results and dashed lines with circle markers correspond to estimation from simulation.}
    \label{fig:phi_vs_tildeqL_appendix}
\end{figure}

In Fig.~\ref{fig:phi_vs_tildeqL_appendix}, we compare the approximate theoretical results $I(\tilde{q}^{L}) \approx M\Phi(\boldsymbol{Q}^*,\boldsymbol{q}^*,...|\tilde{q}^{L})$ to numerical simulations in the scenario of weight sparsification with disconnection probability $p^{l}=1/2$. We observe a good match between the two approaches for sign-DNN, validating the de-correlation assumption of $M$ patterns. For relu-DNN, the theory gives a good prediction on shallow networks with $L=2$ but deteriorates for deeper networks; it suggests the importance of cross-pattern order parameters $q^{l,\perp}$ in this case, whose detailed treatment is beyond the scope of this work.

\section*{References}
\bibliographystyle{iopart-num}
\bibliography{reference}

\end{document}